\documentclass[twocolumn, secnumarabic, amssymb, nobibnotes, aps, prd, superscriptaddress, nofootinbib]{revtex4-2}
\usepackage[left=2cm, right=2cm, top=2.5cm, bottom=2.5cm]{geometry}

\newcommand{\tsp}{\textsuperscript}
\newcommand{\tsb}{\textsubscript}

\usepackage{amsfonts}
\usepackage{amssymb}
\usepackage{amsmath}
\usepackage{dcolumn}
\usepackage{physics}
\usepackage{appendix}
\usepackage{theorem}

\usepackage{lipsum}
\usepackage{mathtools}

\usepackage{tabularx}
\usepackage{multirow}

\usepackage{hyperref}
\hypersetup{
    colorlinks=true,
    linkcolor=blue,     
    citecolor=blue,     
    urlcolor=blue,      
}

\usepackage[version=4]{mhchem}

\usepackage{pythonhighlight}

\usepackage{chemfig}
\usepackage{qcircuit}

\newcommand\arcbetweennodes[3]{%
\pgfmathanglebetweenpoints{\pgfpointanchor{#1}{center}}{\pgfpointanchor{#2}{center}}%
\let#3\pgfmathresult}
\newcommand\arclabel[6][black,-stealth,shorten <=1pt,shorten >=1pt]{%
\chemmove{%
\arcbetweennodes{#4}{#3}\anglestart \arcbetweennodes{#4}{#5}\angleend \draw[#1]([shift=(\anglestart:#2)]#4)arc(\anglestart:\angleend:#2); \pgfmathparse{(\anglestart+\angleend)/2}\let\anglestart\pgfmathresult \node[shift=(\anglestart:#2+1pt)#4,anchor=\anglestart+180,inner sep=0pt,
outer sep=0pt]at(#4){#6};}}
\newcommand\namebond[5][-1pt]{\chemmove{\path(#2)--(#3)node[midway,#4,yshift=#1,black]{#5};}}

\let\oldAA\AA
\renewcommand{\AA}{\text{\normalfont\oldAA}}

\usepackage{graphicx} 
\usepackage[justification=raggedright]{caption}  
\usepackage{subcaption}
\usepackage{soul, xcolor}

\newcommand{\vtheta}{\boldsymbol{\theta}}
\newcommand{\vDelta}{\boldsymbol{\Delta}}

\usepackage{mdframed} 

\newmdenv[
  backgroundcolor=pink,
  linecolor=pink,
  skipabove=\topsep,
  skipbelow=\topsep,
  innertopmargin=0pt,
  innerbottommargin=0pt,
  leftmargin=0pt,
  rightmargin=0pt,
  innerleftmargin=0pt,
  innerrightmargin=0pt,
]{coloreddescription}

\usepackage[normalem]{ulem}
\usepackage{natbib}
\begin{document}

\author{Nacer Eddine Belaloui}
\email{nacer.belaloui@cqtech.org}
\affiliation{Constantine Quantum Technologies, \\Fr\`{e}res Mentouri University Constantine 1, Ain El Bey Road, Constantine, 25017, Algeria}
\affiliation{Laboratoire de Physique Math\'{e}matique et Subatomique, Fr\`{e}res Mentouri University Constantine 1, Ain El Bey Road, Constantine, 25017, Algeria}%
\author{Abdellah Tounsi}
\affiliation{Constantine Quantum Technologies, \\Fr\`{e}res Mentouri University Constantine 1, Ain El Bey Road, Constantine, 25017, Algeria}
\affiliation{Laboratoire de Physique Math\'{e}matique et Subatomique, Fr\`{e}res Mentouri University Constantine 1, Ain El Bey Road, Constantine, 25017, Algeria}%
\author{Rabah Abdelmouheymen Khamadja}
\affiliation{Constantine Quantum Technologies, \\Fr\`{e}res Mentouri University Constantine 1, Ain El Bey Road, Constantine, 25017, Algeria}
\affiliation{Laboratoire de Physique Math\'{e}matique et Subatomique, Fr\`{e}res Mentouri University Constantine 1, Ain El Bey Road, Constantine, 25017, Algeria}%
\author{Mohamed Messaoud Louamri}
\affiliation{Constantine Quantum Technologies, \\Fr\`{e}res Mentouri University Constantine 1, Ain El Bey Road, Constantine, 25017, Algeria}
\affiliation{Theoretical Physics Laboratory, University of
Science and Technology Houari Boumediene, BP 32 Bab Ezzouar,
Algiers, 16111, Algeria}
\author{Achour Benslama}
\affiliation{Constantine Quantum Technologies, \\Fr\`{e}res Mentouri University Constantine 1, Ain El Bey Road, Constantine, 25017, Algeria}
\affiliation{Laboratoire de Physique Math\'{e}matique et Subatomique, Fr\`{e}res Mentouri University Constantine 1, Ain El Bey Road, Constantine, 25017, Algeria}%
\author{David E. {Bernal Neira}}
\affiliation{Davidson School of Chemical Engineering, Purdue University, 480 Stadium Road, West Lafayette, IN, 47907, USA}
\author{Mohamed Taha Rouabah}
\email{m.taha.rouabah@umc.edu.dz}
\affiliation{Constantine Quantum Technologies, \\Fr\`{e}res Mentouri University Constantine 1, Ain El Bey Road, Constantine, 25017, Algeria}
\affiliation{Laboratoire de Physique Math\'{e}matique et Subatomique, Fr\`{e}res Mentouri University Constantine 1, Ain El Bey Road, Constantine, 25017, Algeria}%

\title{
    Ground State Energy Estimation on Current Quantum Hardware Through The Variational Quantum Eigensolver: A Comprehensive Study 
}

\begin{abstract}
    While numerical simulations are presented in most papers introducing new methods to enhance the VQE performance, comprehensive, comparative, and applied studies remain relatively rare. We present a comprehensive, yet concise guide for the implementation of the VQE for molecular problems on NISQ devices, specifically applied to estimate the ground state energy of the \ce{BeH2} molecule using hardware-efficient and chemically informed ansätze. This work addresses several key gaps
in the literature, such as the construction of the electronic Hamiltonian, the transformation of fermionic operators into qubit operators via second quantization, and the mathematical framework's details for the unitary coupled cluster single and double (UCCSD) ansatz. Our methodology, implemented using \texttt{Qiskit 1.2}, the latest release as of the date of this writing, is demonstrated on a noiseless simulator and further tested with noisy quantum circuits. The resilience of the VQE to quantum noise remains an open question. This study compares the computational accuracy of ground state energy estimations for molecules using the VQE across three different current quantum hardware noise models. Furthermore, our experiment on IBM's 156-qubit actual quantum computer revealed valuable insights on the real performance of the VQE on current quantum hardware.

\end{abstract}

\maketitle


\section{Introduction}

Accurately modeling the behavior of molecules at the quantum level provides invaluable insight into chemical properties and processes that are essential to a variety of modern industries. Quantum chemistry \cite{Szabo1996, Levine1999, helgaker2000}, which encapsulates the study of molecules and their properties within quantum mechanics, primarily focuses on the computation of electronic structure properties and their contributions to chemical reactions at the atomic level. In this field, electrons and nuclei are ruled by the Schrödinger equation constructed with the molecular Hamiltonian, and their interactions are described by its solution function, often called the wave function. Quantum chemistry's ability to offer fundamental insights into molecular properties, electronic structures, and chemical reactions makes it a crucial tool in various scientific domains, including medicinal chemistry and drug design \cite{cao2018}, materials chemistry \cite{adjiman2021}, computational biology \cite{fedorov2021}, and environmental chemistry \cite{sandler2002}. For these applications, solving problems related to electronic structure, chemical bonding analysis, molecular ground-state energy, molecular dynamics, and reaction energy are essential \cite{Szabo1996, Levine1999, helgaker2000}.
In particular, accurate estimation of the ground-state energy in quantum chemistry is crucial because it provides a fundamental understanding of the most stable configuration of a molecular system. The ground-state energy is the lowest possible energy that a system can occupy, and it directly influences the physical and chemical properties of molecules. Precise knowledge of this energy allows us to predict molecular behavior, design and optimize new materials, control chemical reactions, and develop pharmaceuticals.

Computational quantum chemistry \cite{helgaker2008quantitative, bowler2012methods, tubman2016} has made significant progress in the past century, with \emph{ab initio} methods \cite{Szabo1996, friesner2005ab, lewars2010, Zhang2022} such as the Coupled-Cluster \cite{lewars2010, lyakh2012multireference} and Møller–Plesset \cite{moller1934, Levine1999, cremer2011moller} methods, but also semiempirical \cite{Szabo1996, lewars2010}, Monte-Carlo \cite{Hammond1991, montanaro2015quantum}, Density Matrix Renormalization Group (DMRG) \cite{white1992, baiardi2020density}, Density Functional Theory (DFT) \cite{cramer2004, lewars2010, yu2016perspective, mardirossian2017thirty} and machine learning \cite{dral2020} methods. 
These approaches utilize systematic approximations to achieve a targeted accuracy based on available resources. Although, despite their success, these methods have limitations and challenges.
From an \emph{ab initio} point of view, to reach a sufficient accuracy in chemical computations, sufficiently large basis sets and full configuration interaction (FCI) wave functions are needed \cite{lewars2010, dobrautz2024}. However, a large basis set requires extensive computational resources \cite{dobrautz2024}, whereas FCI calculations imply a combinatorial complexity that results in a prohibitive computational cost when the number of electrons and the size of the basis set are large. For instance, to the moment of writing this paper, the largest FCI computation has been performed for the $\ce{C3H8}$ molecule within the  Slater-type orbital (STO) 3 Gaussian, \emph{STO-3G}, basis set. The computation included $1.3\times 10^{12}$ configurations and required 256 servers \cite{gao2024}. 
Although computational methods like the DFT give useful predictions and consider electron correlations, they are still approximate and expensive methods for addressing large problems while aiming for an optimal accuracy.
Moreover, as the size of the molecular system grows, the resources required to simulate it classically increase exponentially. That is, quantum chemistry problems can be computationally demanding, especially for large molecules or complex systems \cite{troyer2005, dral2020, gao2024}. 
Furthermore, the interaction between electrons poses a significant challenge for accurately predicting molecular properties, especially in systems with strong electron-electron interactions, which give rise to highly entangled electron states. These states quickly become intractable using \emph{classical} computers since they require heavy FCI computations.
Consequently, accurately solving the Schrödinger equation requires significant computational resources, and approximations become necessary, with the cost of sacrificing precision, to scale quantum computations on classical devices. 

Quantum computers \cite{benioff1980, feynman1982} can overcome some challenges that classical computers face in this matter, and in particular the combinatorial growth of the configuration space of such quantum systems and the correlations within that space \cite{troyer2005}. Indeed, quantum computers would require fewer approximations due to their qubits' ability to be entangled and manifest other quantum behaviors inherently. These behaviors emerge without the need for additional explicit tracking of correlations and probabilities.  By achieving the ability to address these challenges within a feasible time frame without relying extensively on approximations, quantum computers will ultimately enable us to make more accurate predictions regarding various properties of quantum systems \cite{motta2022}.

From the inception of the idea of quantum computers \cite{benioff1980, feynman1982}, through their realizability criteria \cite{divincenzo2000physical}, to current quantum computers \cite{kim2023, wintersperger2023, strohm2024}, numerous algorithms were born in the pursuit of quantum supremacy \cite{motta2022, kim2023}.
An example of the aforementioned algorithms that can be applied to quantum chemistry is the Quantum Phase Estimation (QPE) algorithm \cite{kitaev1995, Nielsen2010}, which enables the computation of all energy levels of a given Hamiltonian while having a polynomial complexity \cite{kitaev1995, Lee2023}, and would have been of a major benefit to the field if it were not for its intolerance to faults. Implementation of QPE requires a fault-tolerant quantum computer with long coherence times and good qubit connectivity, which is still a challenge for all but the smallest of problems, such as the hydrogen molecule which is solvable using only two qubits \cite{Yamamoto2024}. Currently, efforts are more oriented towards utilizing the current \emph{Noisy Intermediate-Scale Quantum} (NISQ) computers, i.e., quantum computers comprised of relatively small numbers (and up to a few hundred) of noisy qubits with short coherence times \cite{preskill2018quantum}.

Variational Quantum Algorithms (VQAs) \cite{Cerezo2021, Scriva2024} have garnered significant attention as promising candidates to achieve quantum advantage with NISQ computers \cite{preskill2018quantum}. VQAs have been developed for a broad spectrum of applications, including the determination of molecular ground states, the simulation of quantum system dynamics \cite{lloyd1996, motta2022}, the solution of linear systems of equations \cite{harrow2009}, and the solution of discrete optimization problems \cite{farhi2014quantum}. VQAs share a unified framework in which tasks are encoded into parameterized cost functions that are evaluated using a quantum computer. A classical optimizer subsequently trains the parameters within the VQA. The inherent adaptability of VQAs makes them particularly well-suited to address the limitations posed by near-term quantum computing technologies.
In this pursuit, \citet{peruzzo2014variational} introduced the Variational Quantum Eigensolver (VQE), a hybrid quantum-classical algorithm that approximates the lowest eigenvalue of a given Hamiltonian. The performance of the VQE algorithm depends on the quality of many components, such as the choice of the ansatz quantum circuit \cite{Tilly2022} and the classical optimizer \cite{pellow2021, sorourifar2024}. The fundamental difference between QPE and VQE is that the former requires the implementation of $O(1)$  quantum circuits with depth $O(1/\epsilon)$ to achieve an energy accuracy of $\epsilon$. In contrast, the latter requires the implementation of $O(1/\epsilon^2)$  quantum circuits with depth $O(1)$ at each iteration \cite{wang2019, Cerezo2021, Scriva2024}.
Nevertheless, the VQE faces significant challenges that limit its current advantages over classical methods for certain applications. One major bottleneck is the high cost of measuring the expectation value of the Hamiltonian, requiring a large number of measurements, especially for complex systems. Although research into efficient operator sampling and parallelization offers potential solutions, these approaches would necessitate a paradigm shift in quantum hardware design. Another limitation lies in the optimization process, which is inherently NP-hard \cite{bittel2021}, with convergence depending on the specific problem's optimization landscape and the choice of optimizer.
Additionally, the presence of barren plateaus in this landscape, where the gradients of the cost function vanish exponentially with system size \cite{larocca2024}, poses severe scaling issues, making optimization intractable for certain parameterizations. Although mitigation strategies, such as identity block initialization and local Hamiltonian encoding, have been proposed, their effectiveness for large-scale systems remains an open question. Similarly, while the VQE shows inherent noise resilience due to its variational nature, error mitigation techniques are often required to achieve accurate results on noisy quantum devices. These methods can significantly increase resource demands, and it is unclear whether this trade-off will be manageable for large-scale applications.
Despite its challenges, the VQE has shown success in small-scale implementations and holds promise as one of the earliest practical algorithms for implementation on NISQ devices. However, realizing its full potential requires continuous advancements not only in quantum hardware but also in the theoretical framework, as well as in the efficiency and robustness of the associated software and algorithms. Such progress is essential to ensure that the benefits of approaches like the VQE can be effectively harnessed at the earliest opportunity.
This paper serves as a practical and thorough guide for utilizing the VQE to estimate molecular ground-state energy, specifically focusing on a single molecular geometry. 
Therefore, we preeminently explore all the steps and components of the VQE and investigate chemically inspired and hardware-efficient ansätze involving a small number of qubits. We specifically benefit from the efficiency of the Simultaneous Perturbation Stochastic Approximation (SPSA) optimizer \cite{spall1998} to mitigate the effects of noise on the evaluation of our cost function \cite{pellow2021}.
By dissecting this process, we aim to provide a foundational understanding {of the entire VQE pipeline} that can be extended to multiple molecular configurations, ultimately aiding in the exploration of {ground and excited states energies} and prediction of molecular dynamics.

\begin{figure}[t!]
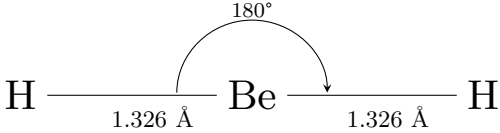

\centering
	\vspace{1cm}
	\setchemfig{bond offset=4pt, atom sep=50pt, atom style={scale=1.75}}
	\chemfig{@{hl}H-@{be}Be-@{hr}H}
	\namebond{hl}{be}{below}{$1.326$ $\AA$ \quad}
	\namebond{be}{hr}{below}{$1.326$ $\AA$}
	\arclabel{1cm}{hl}{be}{hr}{\footnotesize 180\textdegree}
	\vspace{0.5cm}
	\caption{
            The experimental equilibrium molecular geometry of \ce{BeH2} \cite{shayesteh2003,NIST_CCCBDB_2022}.
         }
	\label{fig:beh2}
\end{figure}

Furthermore, {to examine the influence of noise on VQE performance—an essential step toward understanding and enhancing its resilience to noise, especially as quantum hardware advances— we compare state-vector noiseless simulation results to those obtained from noisy quantum circuit simulations using noise models of three recent IBM quantum computers with different error rates, namely IBM Strasbourg, IBM Torino, and IBM Fez. Additionally, we implement the algorithm on the actual IBM Fez quantum computer.
For the sake of this study, we use the VQE to estimate the ground state energy of the \ce{BeH2} molecule at a specific bond length near its stable geometry.\\

This paper is organized as follows:  Section \ref{sec:quant_chem} covers the process of building the Hamiltonian, starting with molecular geometry and molecular orbitals, moving through the second quantization, and concluding with the transformation from fermionic to Pauli operators. In Section \ref{sec:vqe}, we discuss the VQE, the classical optimization, and the specific ansätze employed in our study. Section \ref{sec:sim-qpu-results} presents our experiments, detailing our simulation results, and the implementation of our VQE on a real IBM quantum computer using \texttt{Qiskit 1.2}, the latest version of IBM's SDK at the time of completing this work. Further quantum chemistry computations and details are presented in appendices \ref{app:LCAO+basis-set}, \ref{app:second-quantized-hamiltonian}, and \ref{app:H2-integrals}, while codes are provided in Appendix \ref{app:vqe-pipeline}.

The manuscript can be approached in various ways depending on the reader's background and specific needs. For a complete guide to implementing a VQE for electronic structure problems, the recommended reading sequence is as follows: sections and appendices \ref{sec:quant_chem}, \ref{app:LCAO+basis-set}, \ref{app:second-quantized-hamiltonian}, \ref{app:H2-integrals}, \ref{sec:vqe}, \ref{app:vqe-pipeline}, \ref{sec:sim-qpu-results}, and finally \ref{sec:discussion}. If the details of quantum chemistry calculations are not of particular interest, appendices \ref{app:LCAO+basis-set}, \ref{app:second-quantized-hamiltonian}, and \ref{app:H2-integrals} can be skipped from the above sequence.
The authors have taken significant care to ensure that Appendix \ref{app:vqe-pipeline} is self-contained. Readers who are already familiar with Hamiltonian construction and the components of the VQE and are primarily interested in the \texttt{Qiskit 1.2} implementation may proceed directly to Appendix \ref{app:vqe-pipeline} from this point.

\section{Building the Molecular Hamiltonian}
\label{sec:quant_chem}

In the context of molecular problems, we study the dynamics of a molecule that is comprised of a number of nuclei and electrons, all of which are interacting with each other through the Coulomb force. The general molecular Hamiltonian will thus take the following form:
\begin{multline}   
    H_{mol} = - \sum_n \frac{\mathbf{\nabla}_n^2}{2M_n} - \sum_i \frac{\nabla_i^2}{2} - \sum_{m,i} \frac{Z_m}{|\mathbf{R}_m - \mathbf{r}_i|} \\
    + \sum_{m,n>m} \frac{Z_mZ_{n}}{|\mathbf{R}_m - \mathbf{R}_{n}|} + \sum_{i,j>i}\frac{1}{|\mathbf{r}_i - \mathbf{r}_{j}|},
    \label{eq:mol_hamiltonian}
\end{multline}
in the atomic units. For clarity, we use $m,n$ to sum over nuclei, $i,j$ to sum over electrons, and we use $\mathbf{R}$ and $\mathbf{r}$ to represent position vectors of nuclei and electrons respectively.
The first two terms of \eqref{eq:mol_hamiltonian} are the kinetic energy terms of the nuclei and electrons, respectively, while the last three terms describe (in order) the electron-nucleus interactions, nucleus-nucleus interactions, and electron-electron interactions.
This molecular Hamiltonian can be simplified by transforming it into an electronic Hamiltonian, i.e., a problem where we only solve for the dynamics of the electrons. This is achieved using the Born-Oppenheimer approximation \cite{born1927, Szabo1996} on account of  the large difference between the masses of an electron and that of a nucleus, resulting in a noticeable difference between the speed and frequency of their motion
\footnote{This approximation does not hold under the Jahn-Teller effect where the \emph{conical intersection} takes place, and the excited state interacts with the ground state \cite{bunker1998, yarkony1996}.}.
In this approximation, the nuclei's kinetic energy term, $\sum_n \frac{\nabla_n^2}{2M_n}$, tends to zero. In contrast, the nucleus-nucleus repulsion term, $\sum_{m,n>m} \frac{Z_mZ_{n}}{|\mathbf{R}_m - \mathbf{R}_{n}|}$, becomes a constant that can be computed classically.
After the simplification of the initial Hamiltonian, we obtain the following electronic Hamiltonian:
\begin{equation}
\label{eq:1st-quantized-electronic-hamiltonian}
	H_{el} = - \sum_i \frac{\nabla_i^2}{2} - \sum_{m,i} \frac{Z_i}{|\mathbf{R}_m - \mathbf{r}_i|}  +   \sum_{i,j>i}\frac{1}{|\mathbf{r}_i - \mathbf{r}_{j}|},
\end{equation}
which acts on the wave function $\Psi(\mathbf{x}_1, \mathbf{x}_2, \cdots \mathbf{x}_N)$, where $\mathbf{x}_i = (\mathbf{r}_i, s_i)$ describes the spatial position and the spin of the $i$'th electron. 
To solve for the ground state of the Hamiltonian, quantum mechanical approaches such as \emph{ab initio} methods \cite{Szabo1996, lewars2010}, semi-empirical methods \cite{lewars2010, Levine1999}, and DFT-based approaches \cite{lewars2010, cramer2004} are considered. In this work, we follow the \emph{ab initio} approach, which is based on describing the wave function as a linear combination of Slater determinants of the occupied molecular orbitals. Defining an orthonormal set of molecular orbital allows for the representation of the electronic state as a Fock state, which will be practical for the second quantization of the electronic Hamiltonian in the subsequent section.
The expansion of molecular orbitals as a Linear Combination of Atomic Orbitals (LCAO), which in turn are written in a basis set of Gaussian primitives, is illustrated in Appendix \ref{app:LCAO+basis-set}, alongside the motivation for Gaussian expansions and their different types. The Self-Consistent Field (SCF) method is used to find the values of LCAO coefficients and consequently determines the Hartree-Fock reference state.\\

Finding an exact solution of the Schrödinger equation within a given basis set is equivalent to solving the Full Configuration Interaction (FCI) functions, where the wave function of a molecule is expressed as a linear combination of all possible Slater determinants that can be constructed from a given set of molecular orbitals. However, for a number of electrons $N$ and a number of molecular orbitals $M$, the number of possible occupation configurations increases as $2M \choose N$. Therefore, it is more convenient to use a quantum computer to deal with such factorially growing search space \cite{troyer2005} employing a number of qubits on the scale of $O(\log_2(D))$, where $D$ is the number of determinants.
However, the electronic Hamiltonian in the first quantized form, shown in Eq.\eqref{eq:1st-quantized-electronic-hamiltonian}, is not suitable to simulate and solve for on a quantum computer. Therefore, we need to transform the Hamiltonian to the second quantized operators' form, as the latter will require a finite number of qubits and is more easily mapped into quantum gates. The electronic state in the second quantized form will be represented as a Fock state that encodes the occupation state of each molecular spin orbital. Thus, it represents the Slater determinant of the occupied orbitals. 
The quantum computation advantage lies in the ability to store the coefficients of different Slater determinants in a single quantum register. 

\subsection{The Second Quantization of Electronic Hamiltonian}
\label{sec:second-quantized-hamiltonian}

Since the electronic Hamiltonian involves one- and two-body interaction terms, the second-quantized Hamiltonian can be written under the form:
\begin{equation} \label{eq:2nd_quant}
	H_{el} =  \sum_{p,q} h_{pq} a^\dagger_p a_q + \sum_{p,q,r,s} h_{pqrs} a^\dagger_p a^\dagger_q a_r a_s,
\end{equation}
with $a^\dagger$ and $a$ being the electron creation and annihilation operators. The first term thus represents the transitions of single electrons between different orbitals, while the second term corresponds to the simultaneous transitions of electron pairs between different orbitals. The coefficients $h_{pq}$ and $h_{pqrs}$ are the one- and two-electron integrals defined as \cite{helgaker2008quantitative, lewars2010}
\begin{align}
    \label{eq:h_pq}
    h_{pq} 
    &= \int \psi^*_p(\mathbf{x})\left(\frac{-\nabla^2}{2} - \sum_{i} \frac{Z_i}{|\mathbf{R}_i - \mathbf{r}|}\right)\psi_q(\mathbf{x}) d\mathbf{x},\\
    \label{eq:h_pqrs}
    h_{pqrs} &= \int \frac{\psi^*_p(\mathbf{x}_1)  \psi^*_q(\mathbf{x}_2)  \psi_r(\mathbf{x}_1)  \psi_s(\mathbf{x}_2)}{|\mathbf{r}_1 - \mathbf{r}_2|} d\mathbf{x}_1 d\mathbf{x}_2,
\end{align}
where $\psi(\mathbf{x})$ is the molecular spin orbital's wave function, and $\mathbf{x}$ encapsulates both the electron's position and spin, as defined earlier.
See Appendix \ref{app:second-quantized-hamiltonian} for a detailed derivation of equations \eqref{eq:2nd_quant}, \eqref{eq:h_pq} and \eqref{eq:h_pqrs}.\\

For some selected cases, these integrals can be computed analytically or numerically in a reasonable amount of time. This is especially true in the case of the Gaussian expansion of Slater orbitals. 
In Appendix \ref{app:H2-integrals}, we illustrate the analytical integration of $1s$ type orbitals in the case of the $\ce{H2}$ molecule. The computation of $p$ and higher orbital types' integrals are proven to be efficient using different methods such as the Prism algorithm \cite{gill1991} and Prism-derived algorithms \cite{Barca2016}.
\subsection{Fermionic to Pauli Operators Transformation}
\label{sec:mapping}

\begin{table*}[t] 
    \centering
    \begin{tabular}{p{1.5cm}p{2cm}lcccp{1cm}} 
        \hline
        Active orbitals     & Active electrons  & Mapping & Qubits & Hamiltonian terms & Pauli terms & Average weight \\[0.5ex] 
        \hline
        \multirow{4}{*}{3} & \multirow{4}{*}{2 or 4}& Parity  & 6 & 91  & 34 & 3.12\\
                            && Parity (2 qubit tapered)       & 4 & 91  & 28 & 2.57\\
                            && Jordan-Wigner                  & 6 & 91  & 34 & 2.71\\
                            && Bravyi-Kitaev                  & 6 & 91  & 34 & 3.24\\
        \hline
        \multirow{4}{*}{7} & \multirow{4}{*}{6}& Parity   & 14 & 1939  & 666 & 6.12\\
                            && Parity (2 qubit tapered)   & 12 & 1939  & 666 & 5.69\\
                            && Jordan-Wigner              & 14 & 1939  & 666 & 5.82\\
                            && Bravyi-Kitaev              & 14 & 1939  & 666 & 5.96\\
        \hline
    \end{tabular}
    \caption{In the \ce{BeH2} case, all three transformations require the same number of qubits and produce the same number of terms. The Parity mapping allows for a two-qubit reduction due to the introduced $\mathbb{Z}_2$ symmetry due to the conserved number of $\alpha$ and $\beta$ electrons. In addition, the average weight of the Pauli terms is computed for each method, which indicates the average number of local Pauli measurements required for each term.}
    \label{table:mappings}
\end{table*}

The creation and annihilation operators, $a^\dagger$ and $a$, introduced in the second quantized Hamiltonian \eqref{eq:2nd_quant} are not native to gate-based quantum computers, the latter operating mainly on qubits with Pauli operators. However, a transition from Fermionic to Pauli operators is challenging because it is necessary to maintain the fermionic anti-commutation relations, while single-qubit Pauli operations can only give rise to the bosonic algebra. Several methods have been developed to address this requirement; the most popular include the Jordan-Wigner \cite{Jordan1928}, Parity \cite{Seeley2012}, and Bravyi-Kitaev \cite{bravyi2002} transformations. 
Although the Jordan-Wigner transformation is the natural starting point from an analytical point of view, the successor Parity and Bravyi-Kitaev transformations can be more advantageous. The Parity transformation can introduce a $\mathbb{Z}_2$ symmetry that allows for two-qubit tapering  \cite{bravyi2017, kandala2017}. The Bravyi-Kitaev transformation has the advantage of scaling the weight of Pauli terms, i.e., the number of non-trivial local Pauli operators, logarithmically with the number of qubits instead of linearly. Moreover, different fermionic mapping methods do not, in general, yield the same number of Pauli terms and can differ in measurement performance.
In this work, we use the Parity transformation and qubit tapering since they allow for resource reduction and provide, in this case, lighter Pauli terms that require fewer local measurements, as shown in Table. \ref{table:mappings}.\\

In the Parity transformation, the Fock state is represented as 
\begin{equation}
    \ket{\psi} = \vert e_0 e_1\cdots e_k\rangle,
    \label{eq:fock-state}
\end{equation}
such that $e_i = \left(\sum_{j=0}^{i-1} n_j \mod 2\right)$, where $n_j$ is the occupation number of the orbital $j$ and $e_i$ is the parity of the sum of all occupied orbitals up too the $i$'th, hence the name. Consequently, the ladder operators are given by:
\begin{align}
    a_p = \frac{1}{2} \left(X_pZ_{p-1}-iY_p\right)X_{p+1}\cdots X_{k},
\end{align}
with
\begin{align}
    \{ a_p, a_q \} = \{ a_p^\dagger, a_q^\dagger \}&= 0,\\
    \{ a_p, a_q^\dagger \} &= \delta_{pq}.
\end{align}
In practice, the $\alpha$ and $\beta$ spin sector electrons (spin up and down electrons) can be encoded separately in the Fock state. Notice that the last Pauli operator of $a_p^\dagger a_q$ is either $I_k$ or $Z_k$ and hence commutes with $Z_k$.  Knowing that the number of electrons in the spin up and down sectors is conserved in the electronic Hamiltonian, it is possible to encode $\alpha$ and $\beta$ modes in a bipartite set of qubits. This results in a fixed parity that is encoded in the last qubit of each part. Due to this $\mathbb{Z}_2$ symmetry, it is possible to taper one qubit from each spin sector if the total spin $S^2$ is fixed \emph{a priori} \cite{bravyi2017, kandala2017}. It is worth mentioning that the mapping to Pauli operators is classically efficient since it involves linear relations between ladder and Pauli operators.\\

In the case of \ce{BeH2}, the required number of qubits and Pauli terms is affected by the amount of approximation introduced by fixing the number of active orbitals and electrons as shown in Table \ref{table:mappings}. Such a heavy approximation is not in general recommended. Still, it is necessary in the case of small quantum devices that do not have the required number of qubits for larger Hamiltonians. However, qubit tapering provides a qubit number reduction without introducing approximations by fixing the number of electrons in each of the $\alpha$ and $\beta$ spin sectors.

\begin{figure}[t!]
    \centering
    \includegraphics[width=\linewidth]{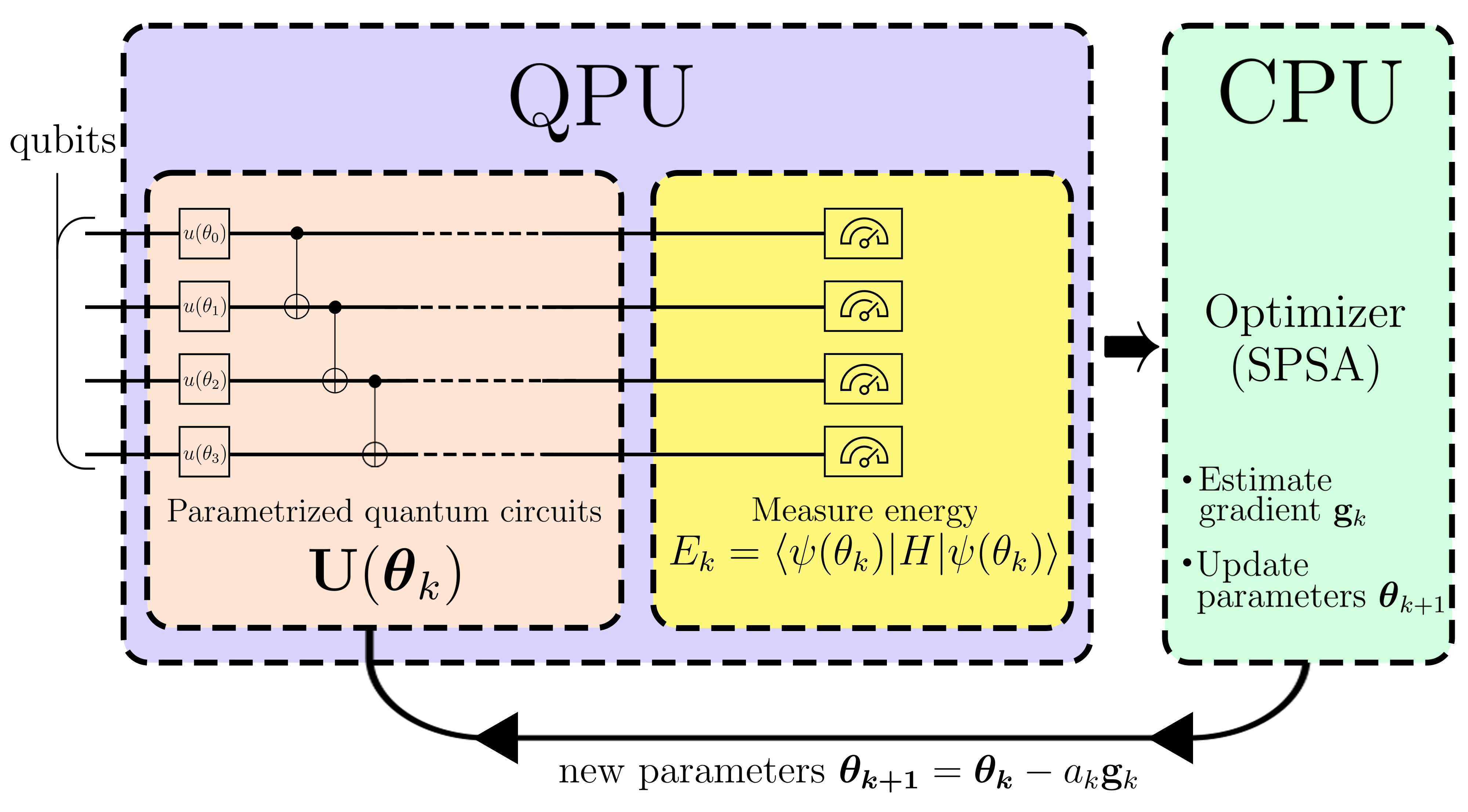}
    \caption{The iterative process and hybrid nature of the VQE. The quantum computer (QPU) is solely used for energy measurements, whereas the classical computer (CPU) is used for parameter optimization. We depict the SPSA as the optimization algorithm.}
    \label{fig:vqe-diagram}
\end{figure}

\section{The Variational Quantum Eigensolver}
\label{sec:vqe}

The variational method in quantum mechanics, and by extension, the variational quantum eigensolver, relies on a trial quantum state to be parametrically adjusted to approximate the exact solution for a given Hamiltonian. The Rayleigh-Ritz theorem~\cite{zettili2009, Tannoudji2017-tome3}, formulated in Eq.\eqref{eq:ritz}, ensures that for any arbitrary trial wave function, the expectation value of the Hermitian Hamiltonian with respect to the trial state is always greater than or equal to the ground state energy, $E_0$, of that Hamiltonian, with closer states to the actual Hamiltonian ground state giving closer expectation values to the ground state energy. Therefore, in the VQE, the trial state should ideally be as physically accurate as possible to obtain accurate results.
Mathematically, the Rayleigh-Ritz theorem for the variational method in quantum mechanics is formulated as follows:
\begin{equation}
    E(\boldsymbol{\theta}) = \frac{\langle \psi(\vtheta)|H| \psi(\vtheta) \rangle}{\braket{\psi(\vtheta)}{\psi(\vtheta)}} \ge E_0,
    \label{eq:ritz}
\end{equation}
with $\vtheta$ being a vector of $n$ real-valued parameters: $\theta_0$, $\theta_1$, $\dots$, $\theta_{n-1}$. And since in our case, $\ket{\psi(\vtheta)}$ is a normalized quantum state that satisfies
\begin{equation}
    \braket{\psi(\vtheta)}{\psi(\vtheta)} = 1,
    \label{eq:normalized-state}
\end{equation}
we can simplify Eq.\eqref{eq:ritz} as
\begin{equation}
    E(\boldsymbol{\theta}) = \langle \psi(\vtheta)|H| \psi(\vtheta) \rangle \ge E_0.
    \label{eq:ritz-simple}
\end{equation}

Making use of Eq.\eqref{eq:ritz}, the VQE process begins by initializing a qubit register. Subsequently, a quantum circuit designed to simulate the physics and entanglements of $\ket{\psi(\vtheta)}$ is applied to this register. We will refer to this quantum circuit as the ansatz.
For the VQE to remain computationally feasible, the circuit depth of the ansatz—the maximum number of quantum gates applied sequentially—must be kept sufficiently low, therefore necessitating the use of a relatively compact ansatz. Once a good ansatz is chosen, the parameters $\vtheta$ are then classically varied iteratively until $E(\boldsymbol{\theta})$ is minimized. Eq.\eqref{eq:ritz} ensures that the minimized energy will converge towards a value that is no lower than the Hamiltonian's ground state energy.\\
The most expensive part of this procedure is the computation of $E(\boldsymbol{\theta})$ given a parameter vector $\vtheta$, especially on a classical computer, as was discussed previously in the introduction. It is thus this computation that will be carried out on a quantum computer. A diagrammatic description of the full VQE procedure is given in Fig.\ref{fig:vqe-diagram}.

\subsection{The Ansatz}
When implementing the VQE for real quantum computers, we face the practical problem of choosing between accurate ansätze and noise-resilient ones.
For quantum chemistry applications, this choice typically lies between the so-called \emph{Hardware-Efficient Ansätze} (\emph{HEA}s) \cite{Tilly2022, Sim2019} that are primarily designed to be implementable on near-term quantum computers, or chemically-inspired ansätze, such as the \emph{Unitary Coupled-Cluster} (\emph{UCC}) ansatz \cite{bakoutsos2018, grimsley2022}. \emph{HEA}s aim to produce high-quality expectation values on noisy quantum computers but may not necessarily be physically informed. This renders the search space they have to cover larger than necessary. On the other hand, chemically-inspired ansätze are designed to model electronic dynamics within the molecule and are thus more suitable for the variational principle under ideal conditions. Still, they are not guaranteed to achieve accurate results on current quantum computers due to their corresponding quantum circuits being deeper. Indeed, NISQ devices are constrained by factors such as noise, limited coherence times, gate fidelity, and qubit connectivity, all of which significantly limit their capability to execute complex or deep quantum circuits reliably. However, it is important to recognize that a shallower ansatz involving fewer quantum operations may lead to reduced accuracy in determining the ground state energy.

\subsubsection{Building a Hardware-Efficient Ansatz}

\begin{figure*}[t!]
    \centering
    \begin{subfigure}[b]{0.49\linewidth}
        \centering
        \includegraphics[width=0.8\linewidth]{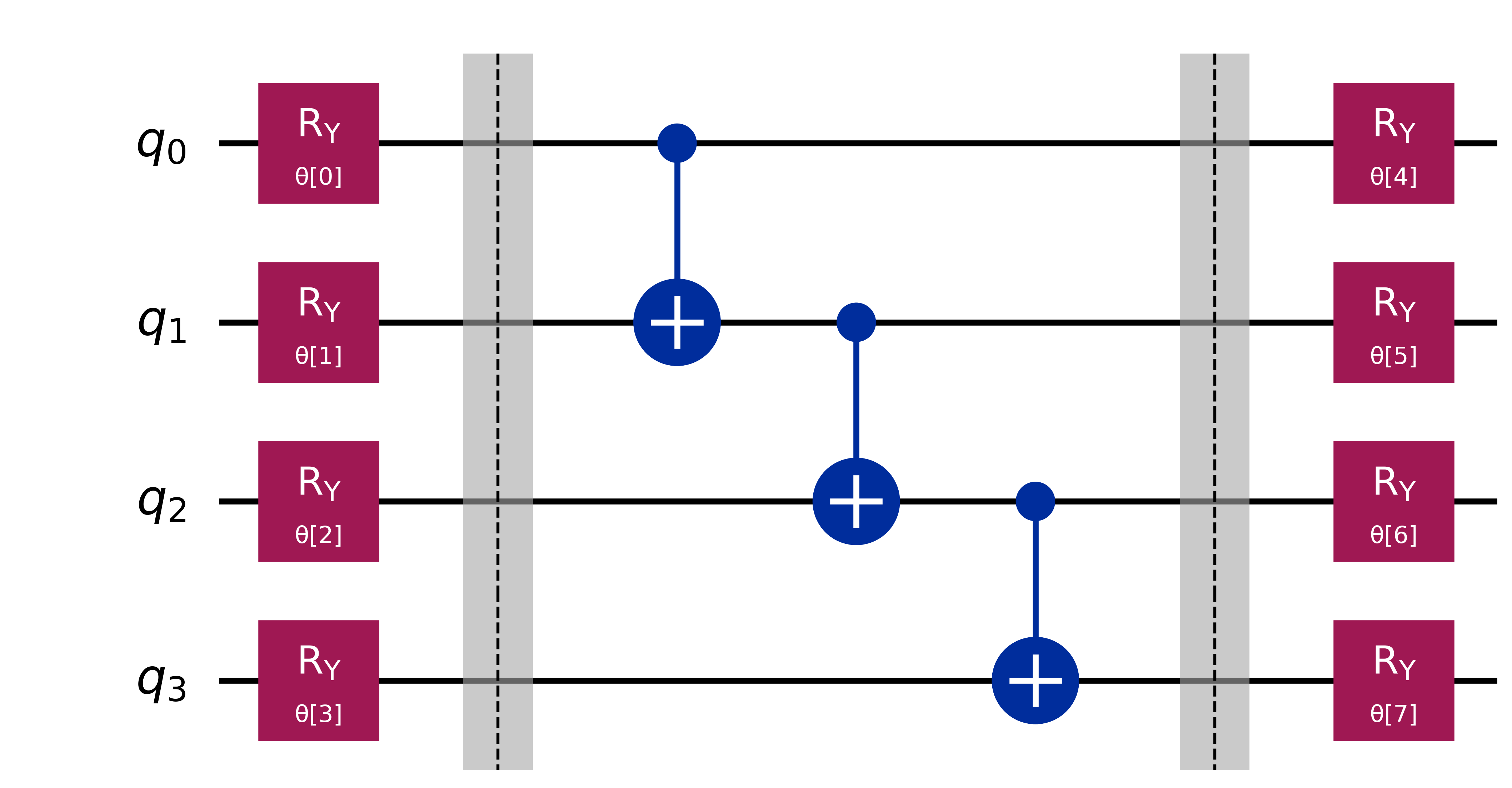}
        \caption{The \emph{Real Amplitudes} ansatz.}
        \label{fig:ra_circuit}
    \end{subfigure}
    \hfill
    \begin{subfigure}[b]{0.49\linewidth}
    \centering
    \includegraphics[width=\linewidth]{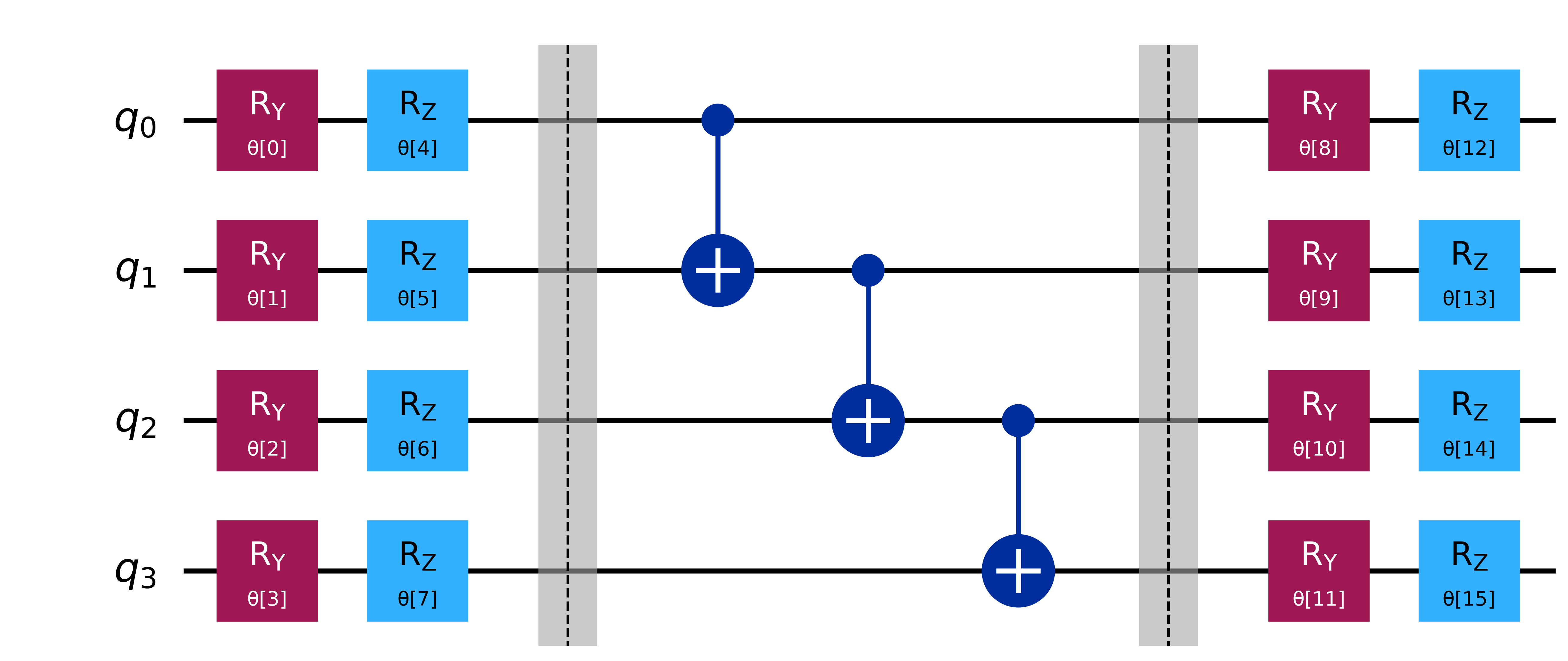}
    \caption{The \emph{Efficient SU2} ansatz.}
    \label{fig:effsu2_circuit}
    \end{subfigure}

    \caption{
        The quantum circuits for two \emph{HEA}s: the \emph{Real Amplitudes} and \emph{Efficient SU2} ansätze. In this example, both ansätze act on four qubits. They are composed of a first rotation gates block, then an entangling block, then a final rotation gates block. The \emph{Efficient SU2} circuit has double the amount of rotation gates since it produces states with complex-valued amplitudes. The circuits were generated and drawn using \texttt{Qiskit} \cite{qiskit}.
    }
    \label{fig:heas}
\end{figure*}

\emph{HEA}s form a broad class of ansätze, which are designed to be usable on near-term quantum computers \cite{Cerezo2021}.
In this approach, unitaries are selected from a set of quantum gates guided by the connectivity and interactions inherent to the target quantum hardware.
This method restrains the circuit's depth increase typically associated with with circuit \emph{transpilation}, where we convert an arbitrary unitary into a sequence of gates that are native to the quantum computer. A key benefit of the hardware-efficient ansätze lies in their adaptability, as they allow for the encoding of symmetries \cite{setia2020} and the closer alignment of correlated qubits for reduced depth, making it particularly advantageous for studying Hamiltonians that closely resemble the device's native interactions \cite{tkachenko2021}.
A widespread construction of \emph{HEA}s is achieved by applying a layer of parameterized rotation gates on all the qubits, followed by a layer of entangling gates also acting on all the qubits \cite{Tilly2022}. These rotations and entangling layer form a block that can be repeated $d$ times. It is common for these circuits to begin and end with the rotation layer. Figures \ref{fig:ra_circuit} and \ref{fig:effsu2_circuit} show two popular \emph{HEA}s: the \emph{Real Amplitudes}, and \emph{Efficient SU2} ansätze, which have moderate expressive and entangling capabilities with a single layer \cite{Tilly2022, Sim2019}. The former requires fewer rotations and parameters and produces quantum states with real coefficients, whereas the latter produces states with complex coefficients at the cost of additional gates and parameters.\\

An important factor to take into account when designing an \emph{HEA} is the qubit connectivity of the target Quantum Processing Unit (QPU). This is due to the fact that entangling, or two-qubit, gates are less accurate than single-qubit gates, and entangling qubits that are not directly connected will require the use of expensive $\operatorname{SWAP}$ gates that will introduce additional noise during the computation \cite{wille2019}. In this work, we adopt the \emph{Efficient SU2} ansatz as our chosen \emph{HEA}; thus, for the remainder of this manuscript, we will refer to it simply as the \emph{Hardware-Efficient Ansatz}, or \emph{HEA}. Fig.\ref{fig:transpiled_effsu2_circuit} shows the difference between the initial logical \emph{HEA} circuit shown in \ref{fig:effsu2_circuit} and the final physical circuit. The latter is executed on the QPU, which in our case is IBM Fez. 

\begin{figure*}[t!]
    \centering
    \includegraphics[width=\linewidth]{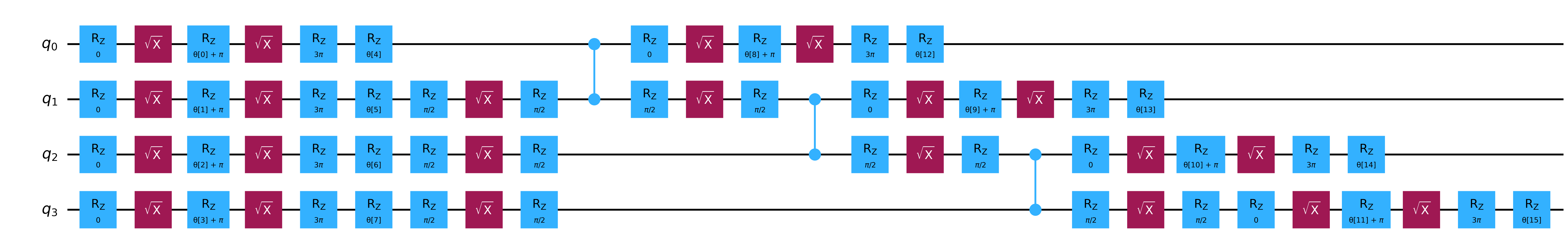}
    \caption{
        The transpiled \emph{HEA} circuit that runs on the IBM Fez QPU. This circuit only uses gates that are physically implemented on the QPU, in this case, the $\operatorname{\sqrt{X}}$, $\operatorname{R_Z}$, and $\operatorname{CZ}$ gates. Therefore, the $\operatorname{R_Y}$ and $\operatorname{CNOT}$ gates have been decomposed. The qubits' mapping has been kept the same since the initial $\operatorname{CNOT}$ gates already follow the target QPU's connectivity.
    }
    \label{fig:transpiled_effsu2_circuit}
\end{figure*}

\subsubsection{Building a Chemically-Inspired Ansatz}

As stated above, an ideal ansatz for the implementation of the VQE in quantum chemistry would model molecular dynamics. A widely used model is the UCC theory, which describes the transitions of electrons from occupied orbitals to unoccupied ones while also modeling their correlations. This can be captured in the following ansatz \cite{watts1989, bakoutsos2018, grimsley2022}:

\begin{equation}
    \ket{\psi(\vtheta)} = e^{\hat{T}(\vtheta) - \hat{T}^\dagger (\vtheta)} \ket{\psi_{\text{init}}},
    \label{eq:ucc_trial}
\end{equation}
where $\hat{T}(\vtheta) = \sum\limits_{i=1}^{n} \hat{T_i} (\vtheta)$ is the cluster operator, which is a sum over $n$ electron excitation operators $\hat{T_i} (\vtheta)$. Each of these operators is written as
\begin{align}
    \hat{T_i}(\vtheta) &= \sum_{\boldsymbol{k}, \boldsymbol{l}} \theta_{k_1 k_2 \cdots k_{i}}^{l_1 l_2 \cdots l_i} a^\dagger_{l_i} \cdots a^\dagger_{l_2} a^\dagger_{l_1} a_{k_i} \cdots a_{k_2} a_{k_1}.
\end{align}
For example, the one- and two-electron excitation operators are
\begin{align}
    \hat{T_1}(\vtheta) &= \sum_{i,j} \theta_{i}^j a^\dagger_j a_i,\\
    \hat{T_2}(\vtheta) &= \sum_{i,j,k,l} \theta_{ij}^{kl} a^\dagger_k a^\dagger_l a_i a_j.
    \label{eq:t1_t2}
\end{align}
$\vtheta$, in this case, is thus the vector of parameters associated with all the possible electron transitions, which are themselves modeled by the creation and annihilation operators $a^\dagger$ and $a$. Since this ansatz preserves the number of electrons, the initial state $\ket{\psi_{\text{init}}}$ is chosen to be one of the possible occupation states, preferably the Hartree-Fock reference state, $\ket{\text{HF}}$.
Because implementing the full \emph{UCC} ansatz is not practical, at least not for near-term quantum computers, as it would require a very deep circuit that implements all excitation operators $\hat{T_i}$, it is common only to consider the \emph{single} and \emph{double} excitation operators $\hat{T_1}$ and $\hat{T_2}$. The resulting restricted ansatz is thus called the \emph{Unitary Coupled Cluster Singles and Doubles} (\emph{UCCSD}) ansatz, where $\hat{T}(\vtheta) \rightarrow \hat{T}_{SD}(\vtheta)$ such as

\begin{align}
    \hat{T}_{SD}(\vtheta) &= \hat{T_1}(\vtheta) + \hat{T_2}(\vtheta)\\
                             &= \sum_{i,j} \theta_{i}^j a^\dagger_j a_i + \sum_{i,j,k,l} \theta_{ij}^{kl} a^\dagger_k a^\dagger_l a_i a_j.
    \label{eq:uccsd_operator}
\end{align}

To implement this \emph{UCCSD} ansatz on a quantum computer, we must go through two essential steps: mapping and a Trotter-Suzuki decomposition, also known as Trotterization, the former having been discussed already in subsection \ref{sec:mapping}. Trotterization is the process of transforming an exponential of a sum of non-commuting operators $\{O_i\}$ into a product of exponentials of single operators \cite{suzuki1976, ikeda2023, avtandilyan2024}:

\begin{equation}
    e^{O_1 + O_2 + \dots} = \lim_{n\to\infty} \left( e^{O_1 / n} e^{O_2 / n} ... \right)^n.
    \label{eq:trotterization}
\end{equation}

The quantum state evolution described in Eq.\eqref{eq:ucc_trial} indeed includes an exponential of a sum of non-commuting operators $\hat{T_i}$ and their adjoints. Explicitly:

\begin{equation}
    \ket{\psi(\vtheta)}_\textit{UCCSD} = e^{\hat{T_1}(\vtheta) + \hat{T_2}(\vtheta) - \hat{T_1}^\dagger(\vtheta) - \hat{T_2}^\dagger(\vtheta)} \ket{\psi_{\text{init}}}.
    \label{eq:uccsd_trial}
\end{equation}

The Trotterization of the evolution operator then gives:

\begin{multline}  
    e^{\hat{T_1}(\vtheta) + \hat{T_2}(\vtheta) - \hat{T_1}^\dagger(\vtheta) - \hat{T_2}^\dagger(\vtheta)} = \\    
    \lim_{n\to\infty} \left( e^{\frac{\hat{T_1}(\vtheta)}{n}} e^{\frac{\hat{T_2}(\vtheta)}{n}} e^{-\frac{\hat{T_1}^\dagger(\vtheta)}{n}} e^{-\frac{\hat{T_2}^\dagger(\vtheta)}{n}} \right)^n.
    \label{eq:uccsd_ev_trott}
\end{multline}

This Trotterization process can present some subtle challenges for near-term quantum computers for two reasons: the first is associated with the exponent $n$ in Eq.\eqref{eq:uccsd_ev_trott}, which should be very large in the exact \emph{UCCSD} solution limit. This means that the circuit simulating the product of exponentials, in our case $( e^{\frac{\hat{T_1}(\vtheta)}{n}} e^{\frac{\hat{T_2}(\vtheta)}{n}} e^{-\frac{\hat{T_1}^\dagger(\vtheta)}{n}} e^{-\frac{\hat{T_2}^\dagger(\vtheta)}{n}} )$, will be repeated $n$ times, for which the execution time may exceed our qubits' coherence time on the one hand, and which leads to an accumulation of noise effects and errors on the other. The second reason is the simulation of each exponential operator, which requires a number of entangling gates that is proportional to the Trotterization degree, $n$, the number of Pauli terms, and their average weight \cite{Nielsen2010, li2021, mukhopadhyay2023}, as shown in Table \ref{table:mappings}.
These two reasons render the implementation of the \emph{UCCSD} ansatz quantum computationally expensive and susceptible to quantum noise and errors. However, it was also numerically shown that in simple molecular systems, a single Trotter step (degree $n=1$) is sufficient for an accurate description of the ground state \cite{romero2019, Tilly2022} since the variational optimization can reduce the effect of the Trotterization error \cite{bakoutsos2018}. We will thus restrict ourselves to a single Trotter step.\\\\
In Table \ref{table:ansatz}, we highlight how the logical entangling gates ($\operatorname{CNOT}$ gates) are decomposed into a greater number of $\operatorname{CZ}$ gates in the transpiled physical circuit corresponding to the utilized quantum computer, further accumulating errors and noise.

\subsection{Optimization}
\label{subsec:optimization}
Varying a set of values in order to minimize a function is a well-known classical procedure termed optimization. It is central to a variety of applications in science, engineering, and machine learning. A plethora of methods and tools for optimization have been developed to be used for a wide range of problems. In this setting, in particular, we are concerned with finding those values of the circuit parameters $\vtheta$ of the ansatz that minimize a cost function. This cost function is the expectation value of the molecular Hamiltonian with respect to the ansatz, and minimizing it corresponds to solving for the Hamiltonian's ground state energy. 
This optimization problem is to be solved using classically implemented algorithms and can be posed as:  
\begin{align}
    \min_{\vtheta}  E(\vtheta) 
 &= \bra{\psi(\vtheta)}  H  \ket{\psi(\vtheta)},
    \label{eq:optimization}
\end{align}
where $\ket{\psi(\vtheta)}$ is the state prepared by the parameterized ansatz, $\vtheta$ is a real-valued parameters vector, and $H$ is the Hamiltonian operator that is to be measured.

After selecting an ansatz, it is crucial to choose a suitable optimizer, as this decision greatly influences both the convergence speed of the VQE optimization process and the overall computational cost of the algorithm, as well as the VQE's resilience to noise in NISQ-era quantum computers. Below is a short description of one such method called the \emph{Simultaneous Perturbation Stochastic Approximation} optimization.

\begin{figure*}[t]
    \centering
    \begin{minipage}[c]{0.39\linewidth}
        \centering
        \subcaptionbox{\label{fig:30_effsu2_svs}}{
        \includegraphics[width=\linewidth]{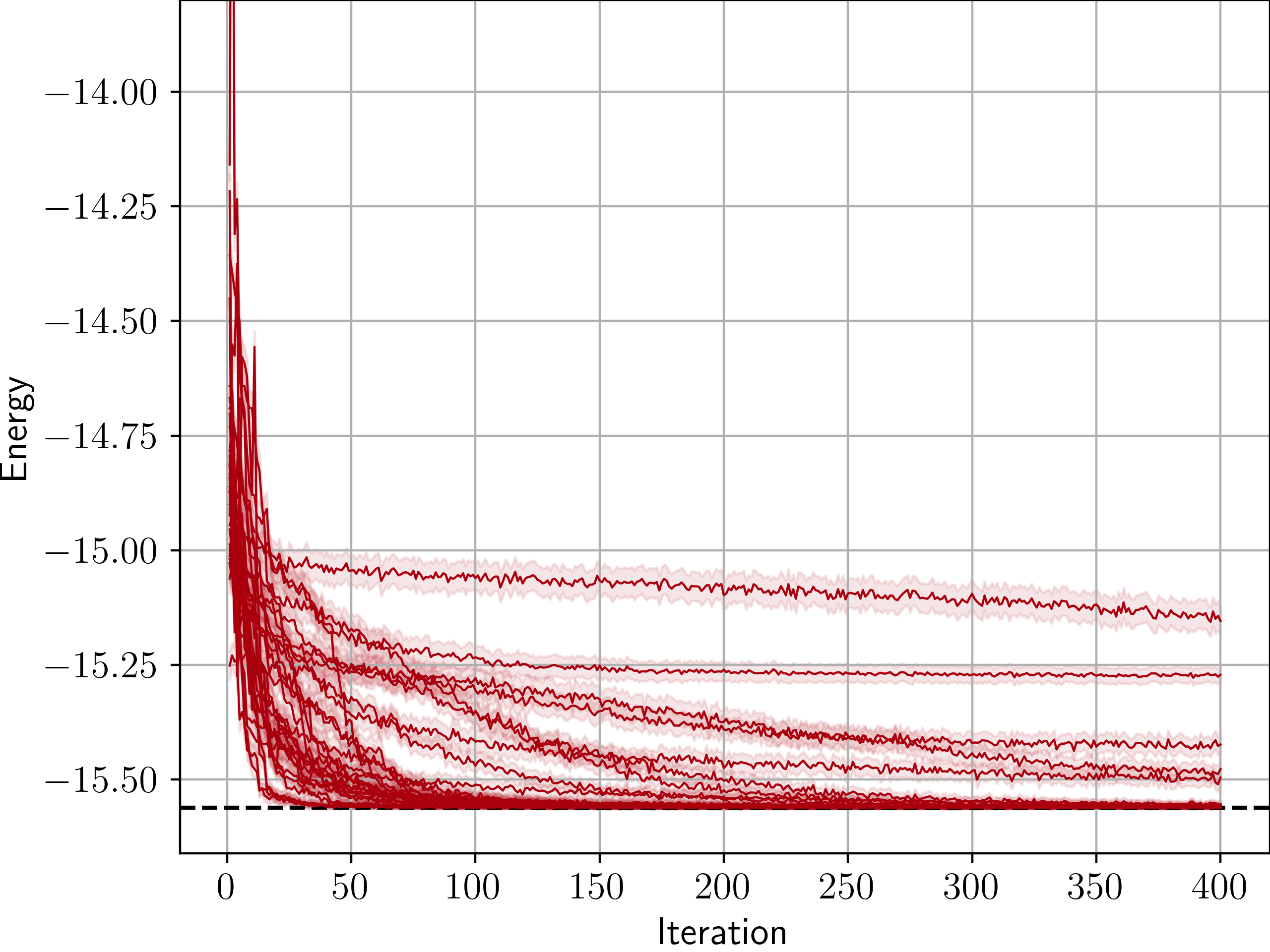}}
    \end{minipage}
    \begin{minipage}[c]{0.39\linewidth}
        \centering
        \subcaptionbox{\label{fig:1.2_uccsd_svs}}{
        \includegraphics[width=\linewidth]{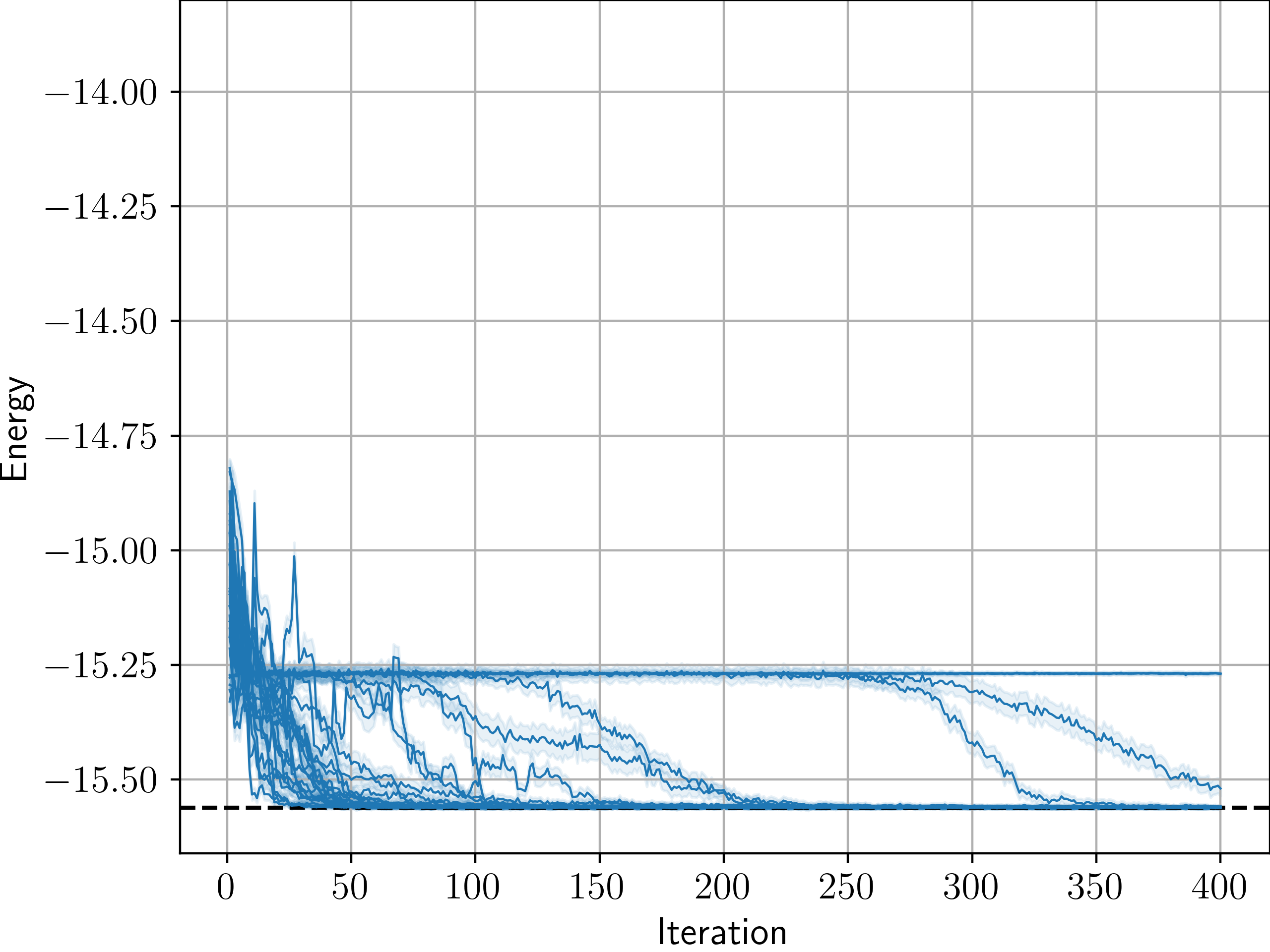}}
    \end{minipage}
    \hfill
    \begin{minipage}[c]{0.19\linewidth}
        \hspace*{0.25ex}
        \includegraphics[width=1\linewidth]{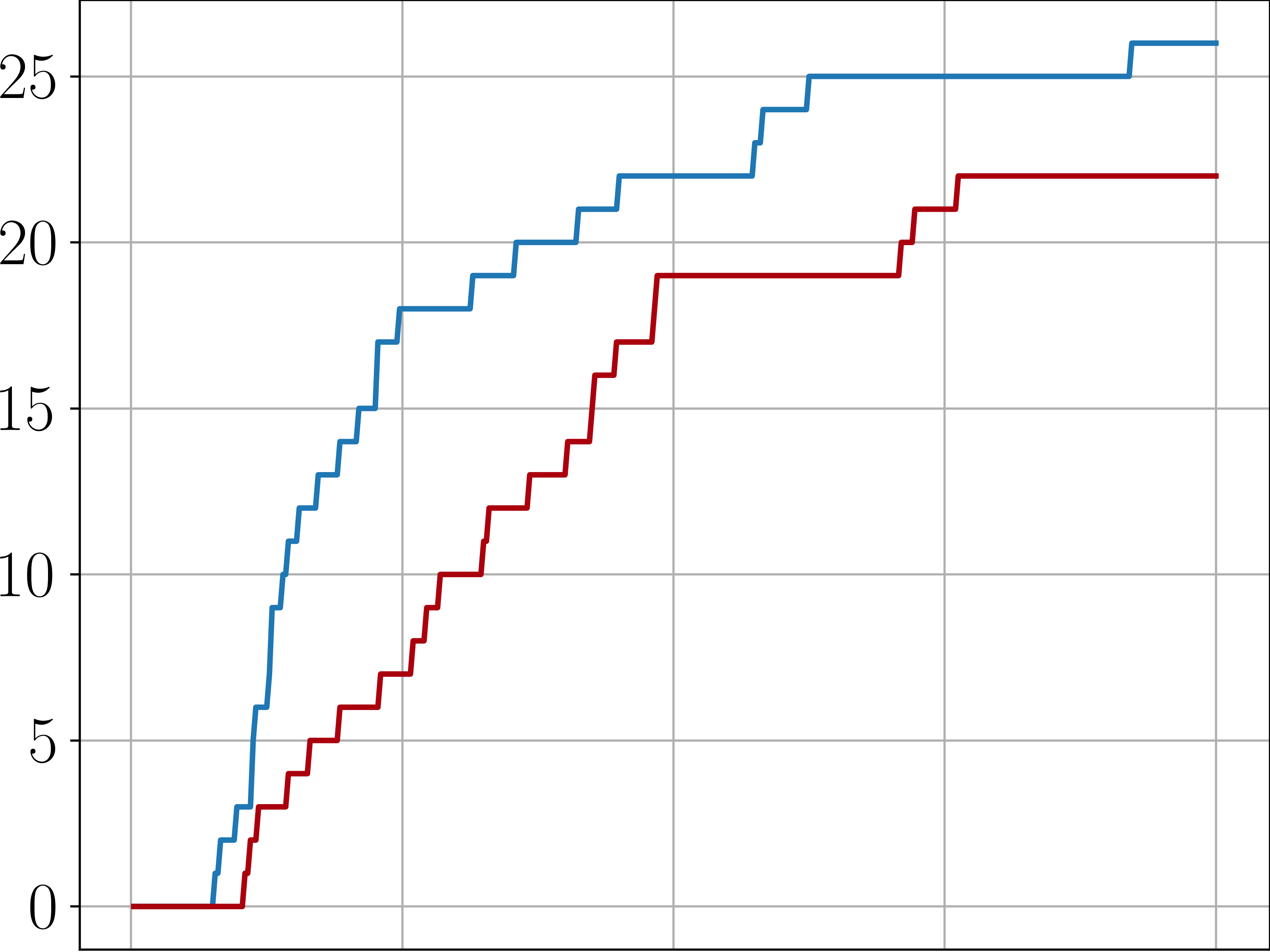}\\
        \subcaptionbox{\label{fig:30_uccsd_svs}}{
        \includegraphics[width=1\linewidth]{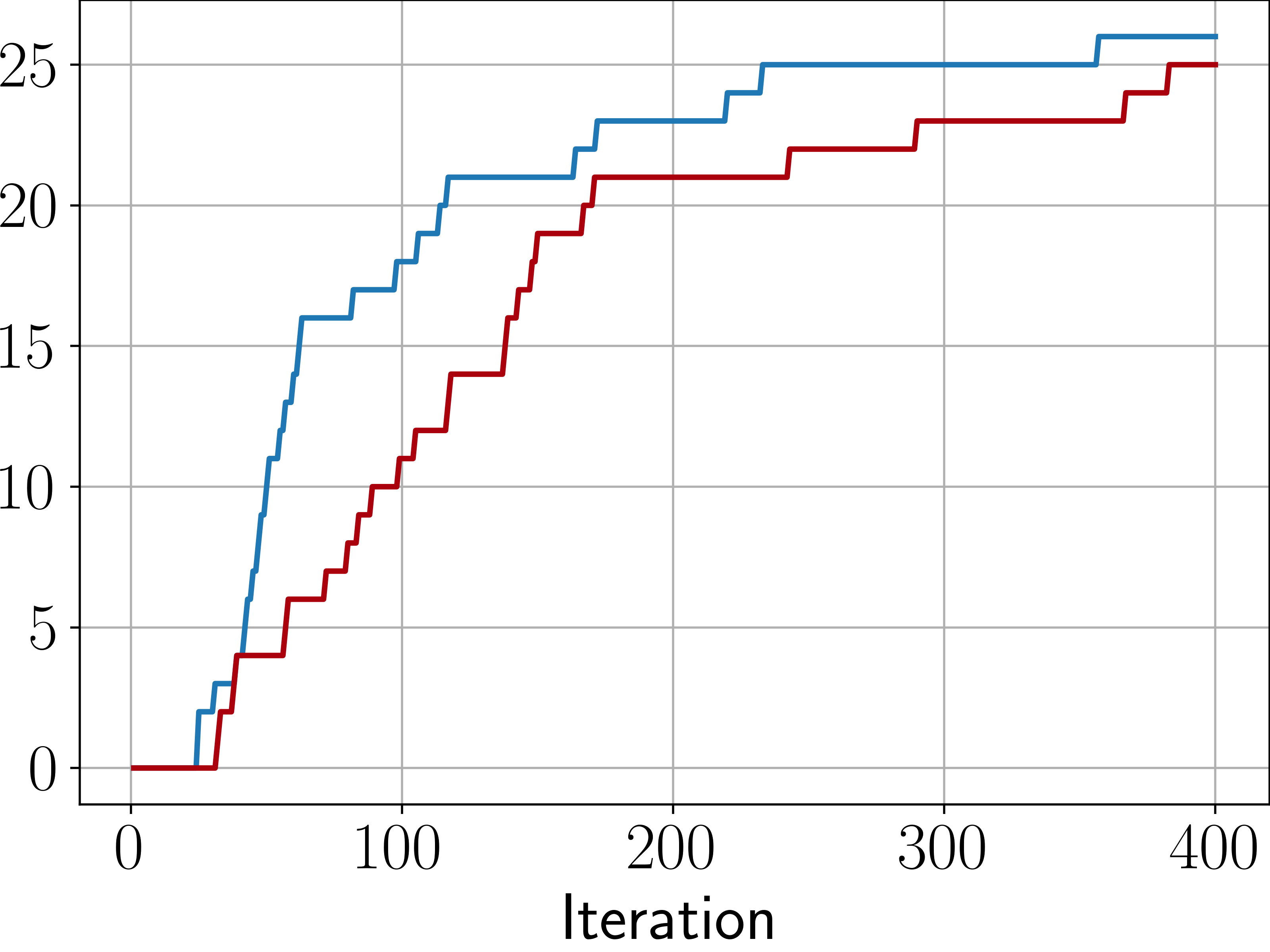}}
    \end{minipage}
    
    \caption{\textbf{(a)} \emph{HEA} (red) and \textbf{(b)} \emph{UCCSD} (blue) simulations on a perfect simulator with shot noise. The shaded area represents the standard deviation of each measurement, which results from 4096 measurement shots. Figures \textbf{(c)} top and bottom show the number of VQEs getting within $1 \times$ and $3 \times 1.6$ mHa of $E^\text{target}$ at each iteration, respectively.}
    \label{fig:all_in_one_with_convergence}
\end{figure*}


\subsubsection*{Simultaneous Perturbation Stochastic Approximation}
The Simultaneous Perturbation Stochastic Approximation (SPSA) is an optimization method that was developed for applications that require optimizing a fluctuating, non-deterministic cost function \cite{spall1998}.
Although initially developed for purely classical applications, it has since proven useful for quantum computing, where it became a popular optimization method due to its performance in powering variational quantum algorithms under noisy conditions \cite{pellow2021}.
The SPSA optimizer requires two energy measurements (cost function calls, in general) $E(\vtheta_k + c_k \vDelta_k)$ and $E(\vtheta_k - c_k \vDelta_k)$ \cite{spall1998} to compute a gradient approximation. The component-wise gradient estimation is thus given by
\begin{equation}
    \mathbf{g}_{k,i}(\vtheta_k) = \frac{E(\vtheta_k + c_k \vDelta_k) - E(\vtheta_k - c_k \vDelta_k)}
                                     {2c_k\vDelta_{k,i}},
    \label{eq:sapa-grad-estimation}
\end{equation}
where $\vtheta_k$ is the vector representing the current set of parameters (at iteration $k$), $\vDelta_k$ is a random vector used to ``perturb" the current parameters $\vtheta_k$, and $c_k$ is a decaying scalar sequence used to attenuate the perturbations as the number of iterations $k$ grows. After the approximate gradient vector $\mathbf{g}_k$ is computed, the next set of parameters is then updated to
\begin{equation}
    \vtheta_{k+1} = \vtheta_{k} - a_k \mathbf{g}_k,
    \label{eq:spsa-params-update}
\end{equation}
where $a_k$ is also a scalar sequence that decays with $k$, called the learning rate. The procedure of estimating $\mathbf{g}_k$ and calculating $\vtheta_{k+1}$ is repeated until the VQE converges towards a minimum of the cost function.
Usually, SPSA starts every optimization with a calibration step, which determines the appropriate learning rate sequence $a_k$ depending on how much the cost function fluctuates. This calibration step requires a number of random cost function evaluations, often set to 50. Finally, $a_k$ and $c_k$ are given by
\begin{equation}
    a_k = \frac{a}{(A + k)^\alpha},
\end{equation}
and
\begin{equation}
    c_k = \frac{c}{k^\gamma},
\end{equation}
where $\alpha$, $\gamma$, $A$, and $c$ are tunable hyperparameters \cite{spall1998, spall2003, kandala2017}.

\section{Simulations and quantum hardware implementation}
\label{sec:sim-qpu-results}

\subsection{Simulations}
\label{subsec:sim}

In this section, we analyze the behavior and convergence of the VQE for \ce{BeH2} under ideal and noisy conditions by means of classical simulations of quantum circuits. We will use the two ansätze we introduced above: the \emph{HEA} and the chemically motivated \emph{UCCSD}.
Classically simulating downscaled versions of the VQE is a good first step to take in order to perform benchmarking and analysis, as well as initial debugging, as classical computing resources are cheaper and easier to access than their quantum counterparts.
The Hamiltonian we are using is that of the \ce{BeH2} at a Be-H bond distance of $1.326 \AA$, with a Complete Active Space (CAS) approximation that includes $2$ electrons and $3$ active molecular orbitals. For further details on how we generate the 2\tsp{nd} quantized Hamiltonian for \ce{BeH2}, refer to the appendix subsection \ref{app:sub:mol-problem}.
In our case, and since we are already dealing with a small-scale VQE, we will be simulating the same quantum circuits to be run on the quantum hardware later.\\

In the following experiments, we perform 30 VQEs for each ansatz and simulator. The initial states associated with the \emph{HEA} and \emph{UCCSD} are $\ket{0000}$ and $\ket{\text{HF}}$, respectively. The initial parameter vectors, $\vtheta$, are randomized for each VQE run. For the parameters optimization, we use SPSA with the following hyperparameters: $\alpha = 0.602$, $\gamma = 0.101$, $A = 0$, and $c=0.2$ \cite{spall2003, kandala2017}.
The optimization procedure starts with an initial $50$ cost function calls to calibrate SPSA's learning rate series $a_k$, while the perturbation series, $c_k$, is determined from the aforementioned hyperparameters. We chose to cut off the optimizer, and therefore the VQE, after $400$ iterations in the simulations. In total, we will thus perform $1251$ measurements: $50$ for the calibration phase, $2 \times 400$ (gradient estimation) $+ \: 400$ (energy measurement) for the optimization, and $1$ final energy measurement.
Lastly, each energy measurement is obtained using $4096$ shots, that is, by measuring every quantum circuit 4096 times and computing the energy from the distribution of measurement results. All simulations ran on \texttt{Qiskit 1.2.0} and \texttt{Qiskit IBM Runtime 0.28.0}. A step-by-step guide to implementing a VQE simulation in \texttt{Qiskit} is provided in Appendix \ref{app:vqe-pipeline}.

\subsubsection{Ideal device simulator}

An \emph{ideal simulator} (or \emph{ideal device}) is an idealized quantum computer that is not affected by any noise channel such as decoherence, gate errors, or readout errors and which is numerically simulated on a classical machine. It may, however, be subject to what is called \emph{shot noise}: fluctuations in the measurements that are due to probabilistic sampling around the classically computed expectation values. This simulates the non-deterministic nature of quantum measurements, even in the idealized case. In this work, the \emph{ideal} VQEs are simulated with shot noise; that is, their measurement results are sampled from a probability distribution. We refer to this idealized device as the state-vector simulator (SVS).
Since we have considered an approximated \ce{BeH2} electronic Hamiltonian, it is worth validating our VQE for both ansätze on an ideal device and analyzing their convergences towards the known ground state energy of the Hamiltonian. The target energy, in this case, is the minimum eigenvalue of the approximated Hamiltonian and is given by
\begin{equation}
     E^{\text{target}} = E_0 = -15.56089 \text{ Ha}. 
    \label{eq:ref_energy}
\end{equation}
We can compute this target value by taking the mapped Hamiltonian in its matrix form and simply diagonalizing it numerically, which gives us the same result as the full configuration interaction (FCI) method. This approach is generally inefficient but is not an issue for our small $4$-qubit Hamiltonian.\\

In figures \ref{fig:best_vqes_by_simulator} and \ref{fig:noisy_to_svs_all}, as well as in the SVS results of Table
\ref{tab:simulations_results}, we observe that both ansätze ended up converging within less than $1$ millihartree (mHa) from the target energy $E^{\text{target}}$. In this ideal case, the \emph{UCCSD} ansatz performs better than the \emph{HEA}. Moreover, as is evident in Fig.\ref{fig:all_in_one_with_convergence}, we notice a faster convergence for the \emph{UCCSD} compared to the \emph{HEA}, with a small number of VQEs converging towards a local minimum situated at around $-15.25$ Ha.
This validates both the \emph{UCCSD} and the \emph{HEA} as potentially good candidates for our molecular problem.

\subsubsection{Noisy simulator}
\begin{figure*}[t!]
    \centering
    \begin{subfigure}[b]{0.49\linewidth}
        \centering
        \includegraphics[width=\linewidth]{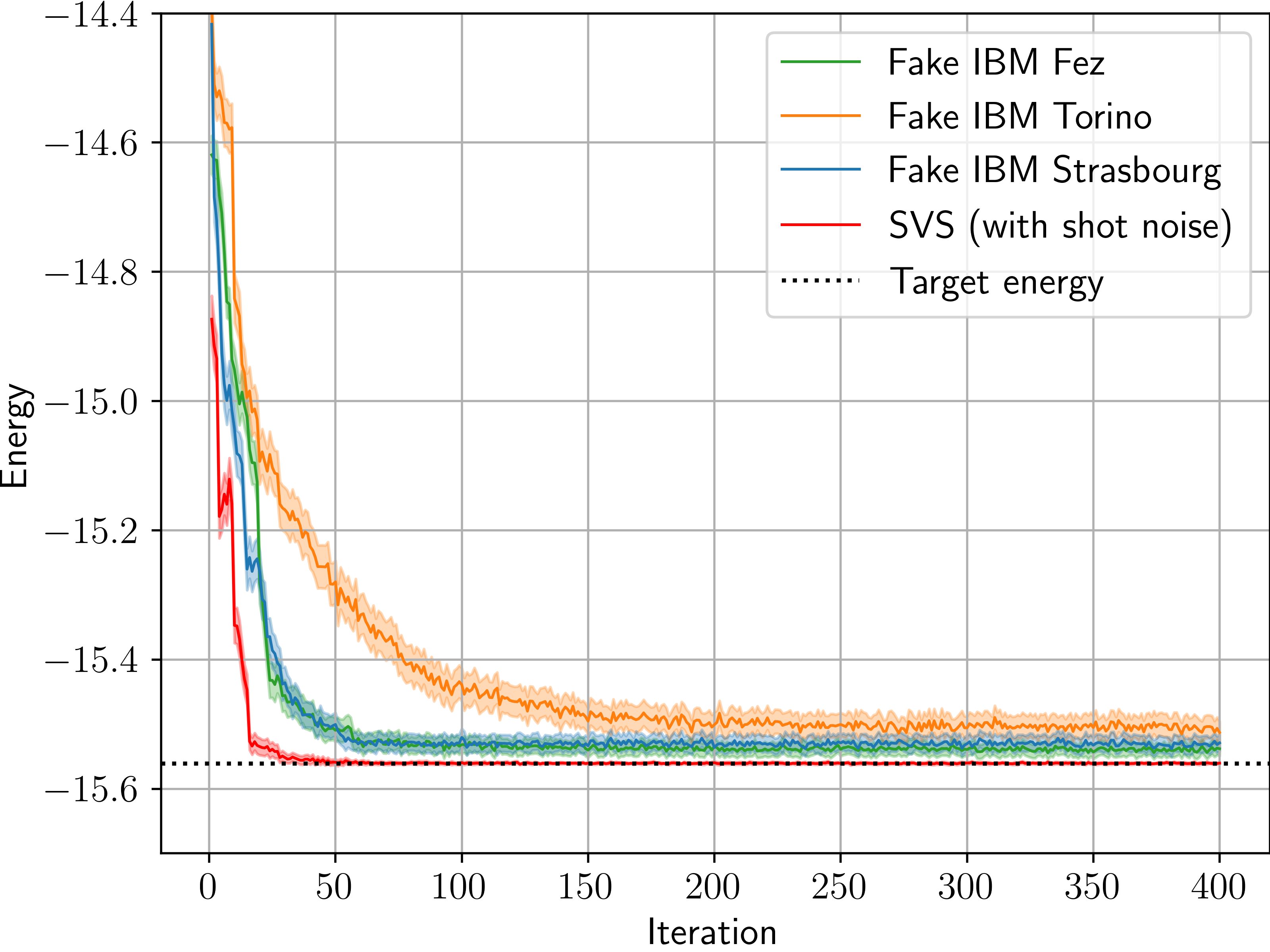}
        \caption{The \emph{HEA} best VQEs by simulator.}
        \label{fig:effsu2_best_vqe_by_sim}
    \end{subfigure}
    \hfill
    \begin{subfigure}[b]{0.49\linewidth}
    \centering
    \includegraphics[width=\linewidth]{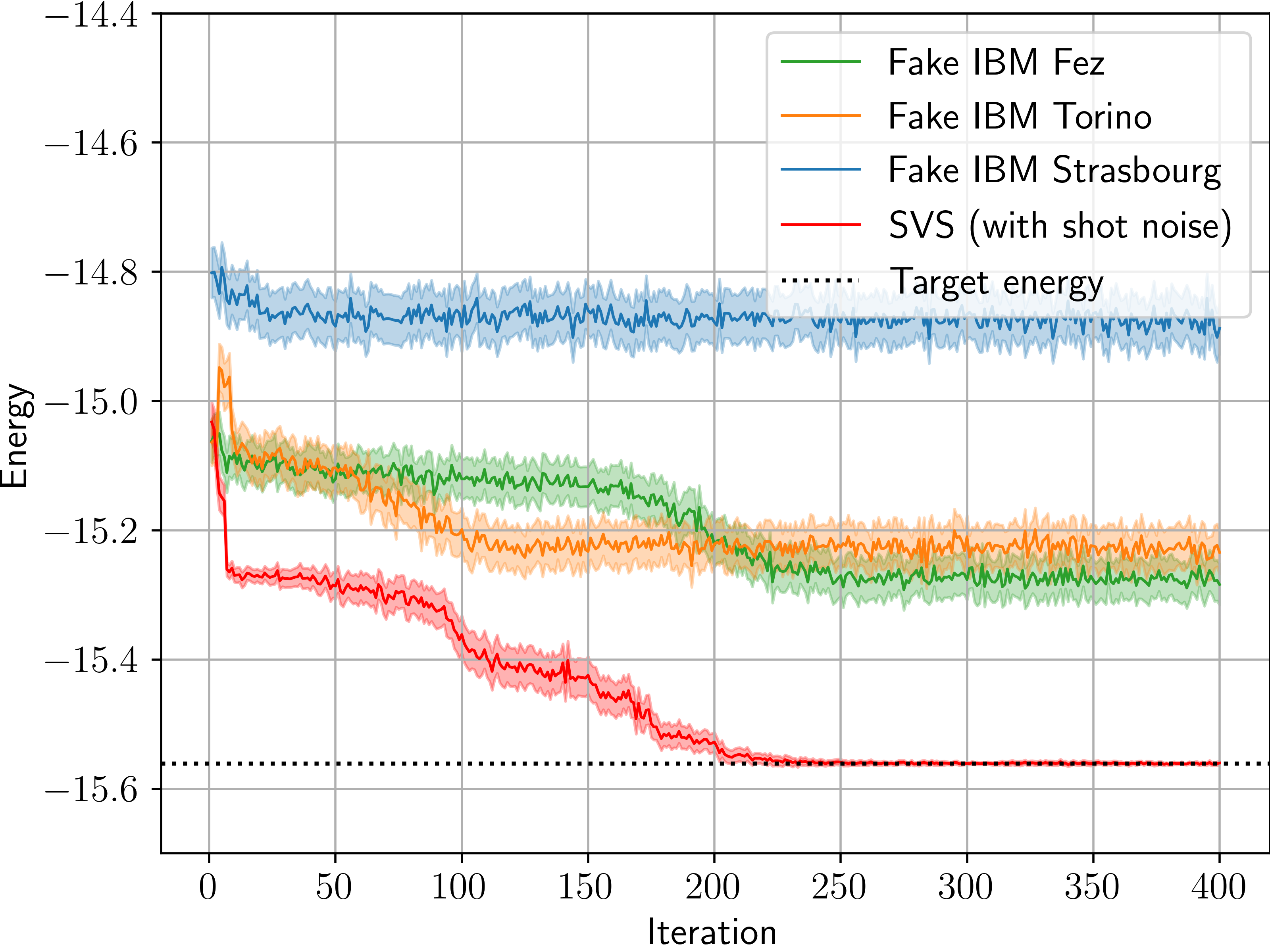}
    \caption{The \emph{UCCSD} best VQEs by simulator.}
    \label{fig:uccsd_best_vqe_by_sim}
    \end{subfigure}

    \caption{The convergence graphs of the best-performing VQE per simulator. For each simulator, the 30 VQEs are sorted by the average of their last 40 energies (last 10\% of iterations). The VQEs with the lowest average are shown here. The shaded areas correspond to the standard deviation of each measurement (from 4096 shots).}
    \label{fig:best_vqes_by_simulator}
\end{figure*}

\begin{figure*}[t]
    \centering
    \begin{subfigure}[b]{0.49\linewidth}
        \centering
        \includegraphics[width=\linewidth]{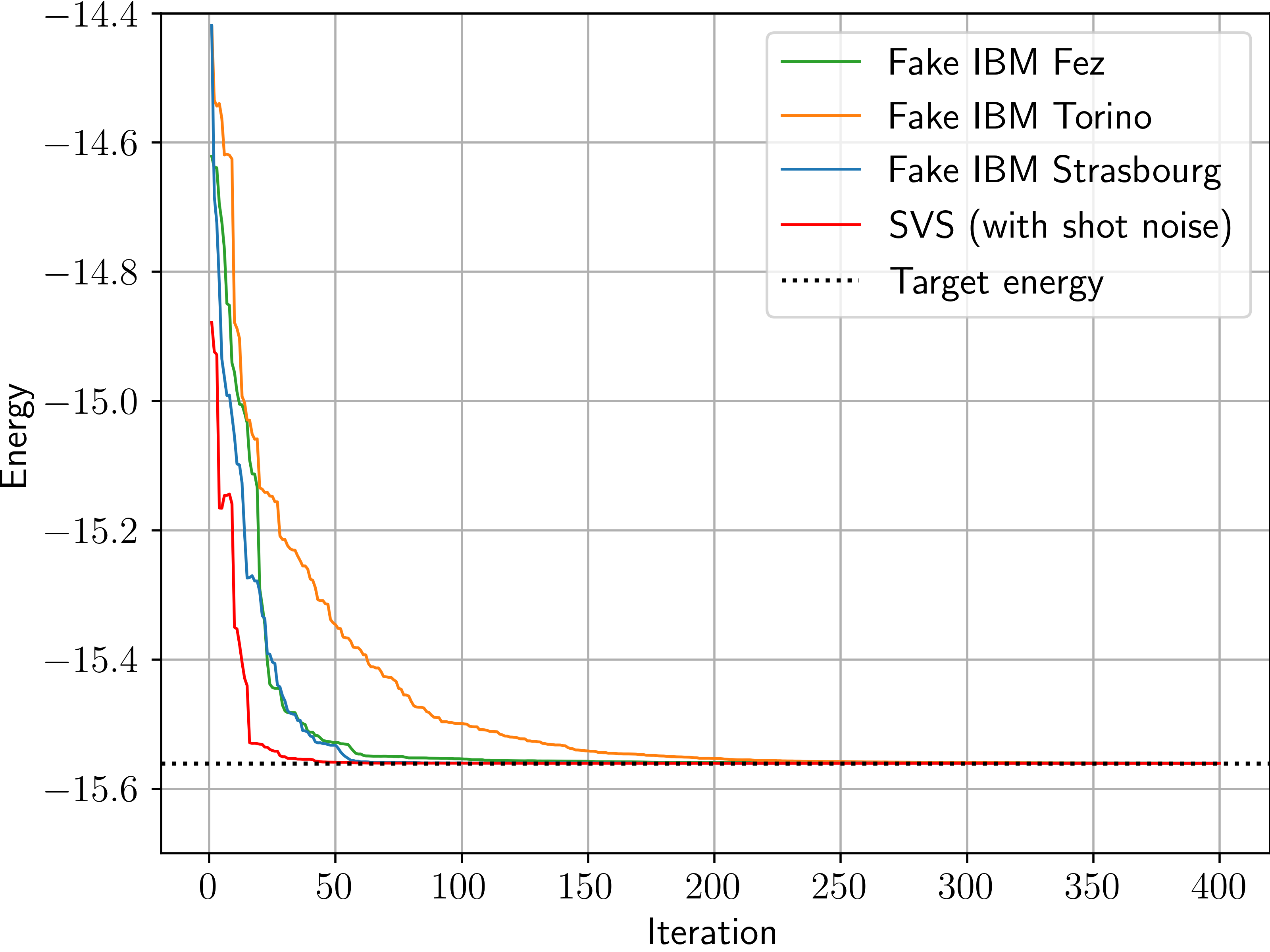}
        \caption{The \emph{HEA} noisy parameters on SVS.}
        \label{fig:effsu2_noisy_to_svs_all}
    \end{subfigure}
    \hfill
    \begin{subfigure}[b]{0.49\linewidth}
    \centering
    \includegraphics[width=\linewidth]{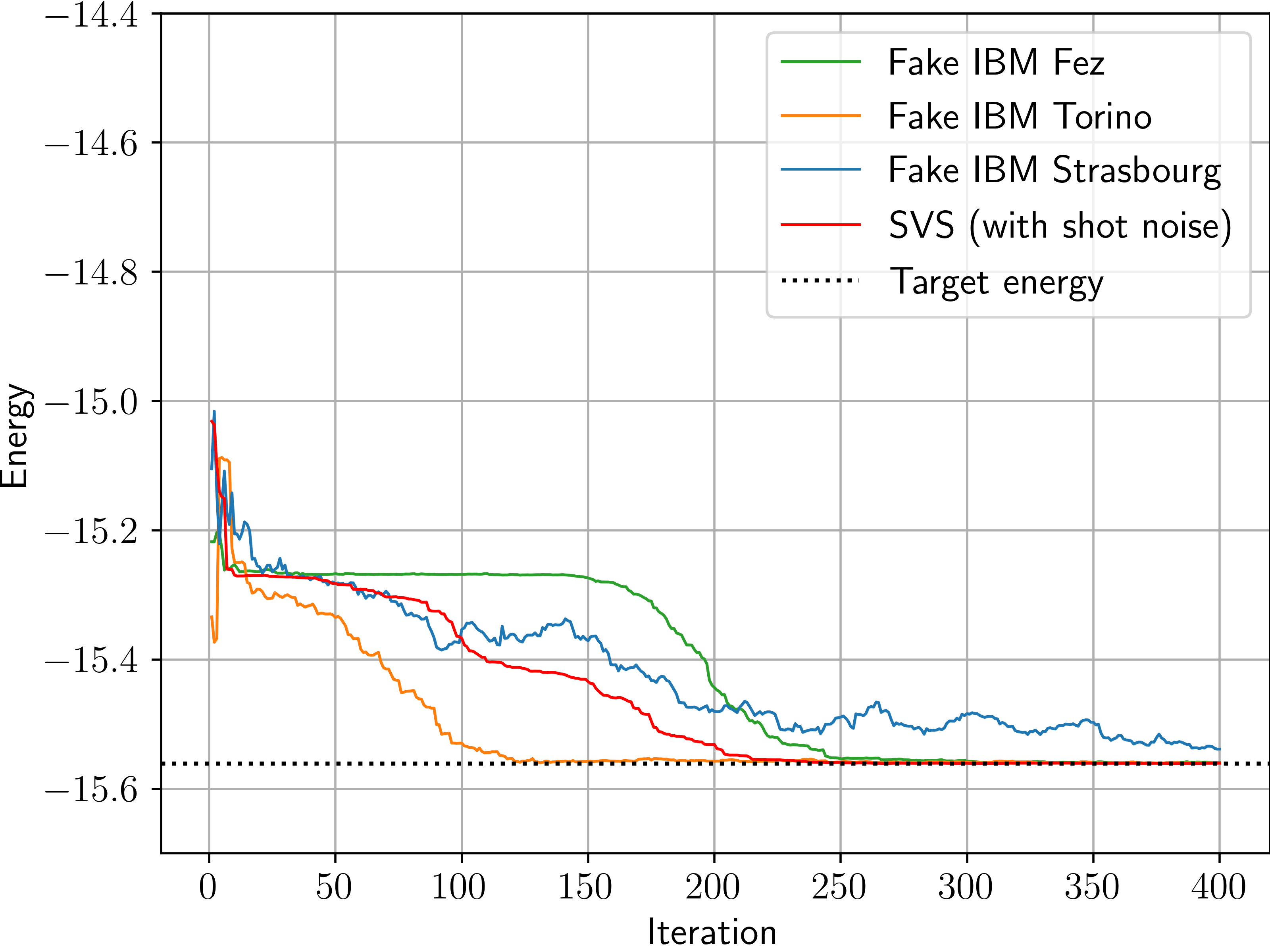}
    \caption{The \emph{UCCSD} ansatz  noisy parameters on SVS.}
    \label{fig:uccsd_noisy_to_svs_all}
    \end{subfigure}

    \caption{Convergence graphs with energies evaluated on SVS with no shot noise using the total set of parameters $\{\vtheta_\text{best}\}$ of the best-performing VQE per ansatz and simulator.}
    \label{fig:noisy_to_svs_all}
\end{figure*}

\begin{table}[t!]
    \centering
    Logical Circuits \\[1.5ex]
    \begin{tabular}{lccc} 
        \hline
        Ansatz                  & Depth & $\operatorname{CNOT}$s & Parameters\\[0.5ex] 
        \hline
        \text{UCCSD}          & 315   & 172   & 8    \\
        \text{HEA}            & 7     & 3     & 16   \\[0.5ex] 
        \hline
    \end{tabular}

    \vspace{8pt}
    
    Transpiled Circuits \\[1.5ex]
    \begin{tabular}{lccc} 
        \hline
        Ansatz                & Depth & $\operatorname{CZ}$s & Parameters\\[0.5ex] 
        \hline
        \text{UCCSD} (unoptimized)      & 1258   & 256   & 8    \\
        \text{UCCSD} (optimized)        & 615    & 185   & 8    \\
        \text{HEA}   (unoptimized)      & 27     & 3     & 16   \\
        \text{HEA}   (optimized)        & 21     & 3     & 16   \\[0.5ex] 
        \hline
    \end{tabular}
    \caption{A comparison of the logical and transpiled \emph{UCCSD} ansatz and the \emph{HEA} circuits to be run on real hardware in terms of depth, number of entangling gates ($\operatorname{CNOT}$ and $\operatorname{CZ}$ gates), and number of parameters. We consider both unoptimized and optimized transpilations with \texttt{Qiskit}'s transpiler's optimization levels $0$ and $3$, respectively. The target real hardware here is the 156-qubit IBM Fez.}
    \label{table:ansatz}
\end{table}

After validating that both ansätze converge within chemical accuracy (less than $1.6 \text{ mHa}$ from $E^{\text{target}}$) in the ideal case, the next step is to investigate their performance when noise is introduced. This is done by simulating the noise profile of the target quantum device, as well as the device's physical characteristics such as the qubits' connectivity and the natively supported quantum gates. We chose to use noise models of the following IBM quantum computers: IBM Fez, IBM Torino, and IBM Strasbourg. The first two are of the Heron family, the latest IBM QPU family as of the time of writing, while the latter is of the precedent Eagle family. Heron QPUs have a higher number of qubits, lower noise levels, and a different 2-qubit entangling gate than Eagle QPUs \cite{IBMQ2024}. We specifically selected these three QPUs to showcase how the VQE results change mainly as a function of noise levels, with the least noisy QPU being IBM Fez, and the most noisy being IBM Strasbourg.\\
Table \ref{table:ansatz} compares the \emph{UCCSD} ansatz and the \emph{HEA} in terms of circuit depth and number of entangling gates as they would be implemented on the QPU, in addition to the number of parameters in each ansatz. In our simulations, we have used \texttt{Qiskit}'s optimization level $3$ to transpile the \emph{UCCSD} circuit. This is used for all the \emph{UCCSD} VQEs. 

\begin{table*}[t!]
    \centering
    \begin{tabular}{c c}
        \begin{minipage}{0.49\textwidth}
            \centering
            \emph{HEA}\\[1.5ex]
            \setlength{\tabcolsep}{10pt} 
            \begin{tabular}{@{\hspace{3pt}}llc@{\hspace{3pt}}}
                \hline
                Simulator                      & $\langle E_{\rm VQE}(\{\vtheta_{\text{best}}\}_{\rm last 10\%})\rangle$  & $\Delta E_{\rm VQE}$  \\ [0.7ex] 
                \hline
                IBM Torino                      & $-15.50513(445)$         &  $0.05575$   \\ [0.5ex]
                IBM Strasbourg                  & $-15.53111(291)$         &  $0.02978$   \\ [0.5ex]
                IBM Fez                         & $-15.53934(243)$         &  $0.02155$   \\ [0.5ex]
                SVS                             & $\mathbf{-15.56055(51)}$     &  $\mathbf{0.00034}$   \\ [0.5ex]
                \hline \\ [-1.5ex]
                $E^{\rm target}$                & $-15.56089$              & $-$          \\ [0.5ex]
                \hline
            \end{tabular}
            
            \vspace{10pt}
            
            \emph{UCCSD}\\[1.5ex]
            \begin{tabular}{@{\hspace{3pt}}llc@{\hspace{3pt}}} 
                \hline
                Simulator                      & $\langle E_{\rm VQE}(\{\vtheta_{\text{best}}\}_{\rm last 10\%})\rangle$  & $\Delta E_{\rm VQE}$ \\ [0.7ex] 
                \hline
                IBM Strasbourg                 & $-14.87717(1304)$        &  $0.68371$   \\ [0.5ex]
                IBM Torino                     & $-15.22978(1002)$        &  $0.33111$   \\ [0.5ex]
                IBM Fez                        & $-15.27463(724)$         &  $0.28625$   \\ [0.5ex]
                SVS                            & $\mathbf{-15.56084(76)}$ &  $\mathbf{0.00005}$   \\ [0.5ex]
                \hline \\ [-1.5ex]
                $E^{\rm target}$                & $-15.56089$              & $-$          \\ [0.5ex]
                \hline
            \end{tabular}
            \caption*{(a)}
        \end{minipage}
        \begin{minipage}{0.49\textwidth}
            \centering
            \emph{HEA on SVS} \\[1.5ex]
            \setlength{\tabcolsep}{10pt} 
            \begin{tabular}{@{\hspace{3pt}}llc@{\hspace{3pt}}} 
                \hline
                Simulator                      & $\langle E_{\rm SVS}(\{\vtheta_{\text{best}}\}_{\rm last 10\%})\rangle$  & $\Delta E_{\rm SVS}$  \\ [0.7ex] 
                \hline
    
                IBM Torino                      & $\mathbf{-15.55968(40)}$          &  $\mathbf{0.00121}$   \\ [0.5ex]
                IBM Fez                         & $\mathbf{-15.55993(14)}$          &  $\mathbf{0.00096}$   \\ [0.5ex]
                IBM Strasbourg                  & $\mathbf{-15.56021(6)}$           &  $\mathbf{0.00067}$   \\ [0.5ex]
                SVS                             & $\mathbf{-15.56053(2)}$           &  $\mathbf{0.00036}$   \\ [0.5ex]
                \hline \\ [-1.5ex]
                $E^{\rm target}$                & $-15.56089$              & $-$          \\ [0.5ex]
                \hline
            \end{tabular}
            
            \vspace{10pt}
            
            \emph{UCCSD on SVS} \\[1.5ex]
            \begin{tabular}{@{\hspace{3pt}}llc@{\hspace{3pt}}} 
                \hline
                Simulator                      & $\langle E_{\rm SVS}(\{\vtheta_{\text{best}}\}_{\rm last 10\%})\rangle$  & $\Delta E_{\rm SVS}$  \\ [0.7ex] 
                \hline
                IBM Strasbourg                 & $-15.51147(1580)$        &  $0.04942$   \\ [0.5ex]
                IBM Fez                        & $-15.55865(136)$         &  $0.00224$   \\ [0.5ex]
                IBM Torino                     & $-15.55916(77)$          &  $0.00173$   \\ [0.5ex]
                SVS                             & $\mathbf{-15.56050(21)}$         &  $\mathbf{0.00039}$   \\ [0.5ex]
                \hline \\ [-1.5ex]
                $E^{\rm target}$                & $-15.56089$             & $-$          \\ [0.5ex]
                \hline
            \end{tabular}
            \caption*{(b)}
        \end{minipage}
    \end{tabular}
    \caption{
        Mean energy value over the last 10\% iterations (40), alongside the standard deviation resulting from this average for (a) each best VQE for each ansatz and simulator, (b) the SVS-evaluated energies of each best VQE. $E^{\rm target}$ corresponds to the exact energy in the limit of the used level of theory, and $\Delta E = |\langle E \rangle - E^{\rm target}|$.
        All energies are in Ha. Values within chemical accuracy ($\Delta E \leq 0.0016$ Ha) are in bold.
    }
    \label{tab:simulations_results}
\end{table*}


Fig.\ref{fig:best_vqes_by_simulator} shows the best noisy simulations graphs and Fig.\ref{fig:noisy_to_svs_all} shows the evaluation of energy values corresponding to the same best parameters on SVS, while in Table \ref{tab:simulations_results} we present the mean energy values over the last 10\% iterations (40 in our case) for the best VQE results obtained on noisy simulators and the SVS energy evaluations of these best noisy results. We define the best result as the VQE with the lowest average energy over its last $10\%$ of iterations. Interestingly, the two ansätze were affected differently by the introduced noise. 
VQEs with both the \emph{UCCSD} and the \emph{HEA} now converge towards a higher energy value compared to the ideal case, with the \emph{HEA} performing significantly better. However, the added noise affected the quality of the resulting optimized ansatz parameters much less than it affected the energy estimations, as is shown by the evaluation of these parameters on the SVS with no shot noise. Again, the \emph{HEA} gave better results in this regard compared to \emph{UCCSD}, although the \emph{UCCSD} resulting optimized parameters are revealed to be much better than what the noisy energy estimates are indicating. 
These findings suggest that the VQE can be somewhat robust to the simulated levels of noise when it comes to parameter optimization, even if the measured energies are inaccurate.

\subsection{QPU Experiment}
\begin{figure}
    \centering
    \includegraphics[width=0.95\linewidth]{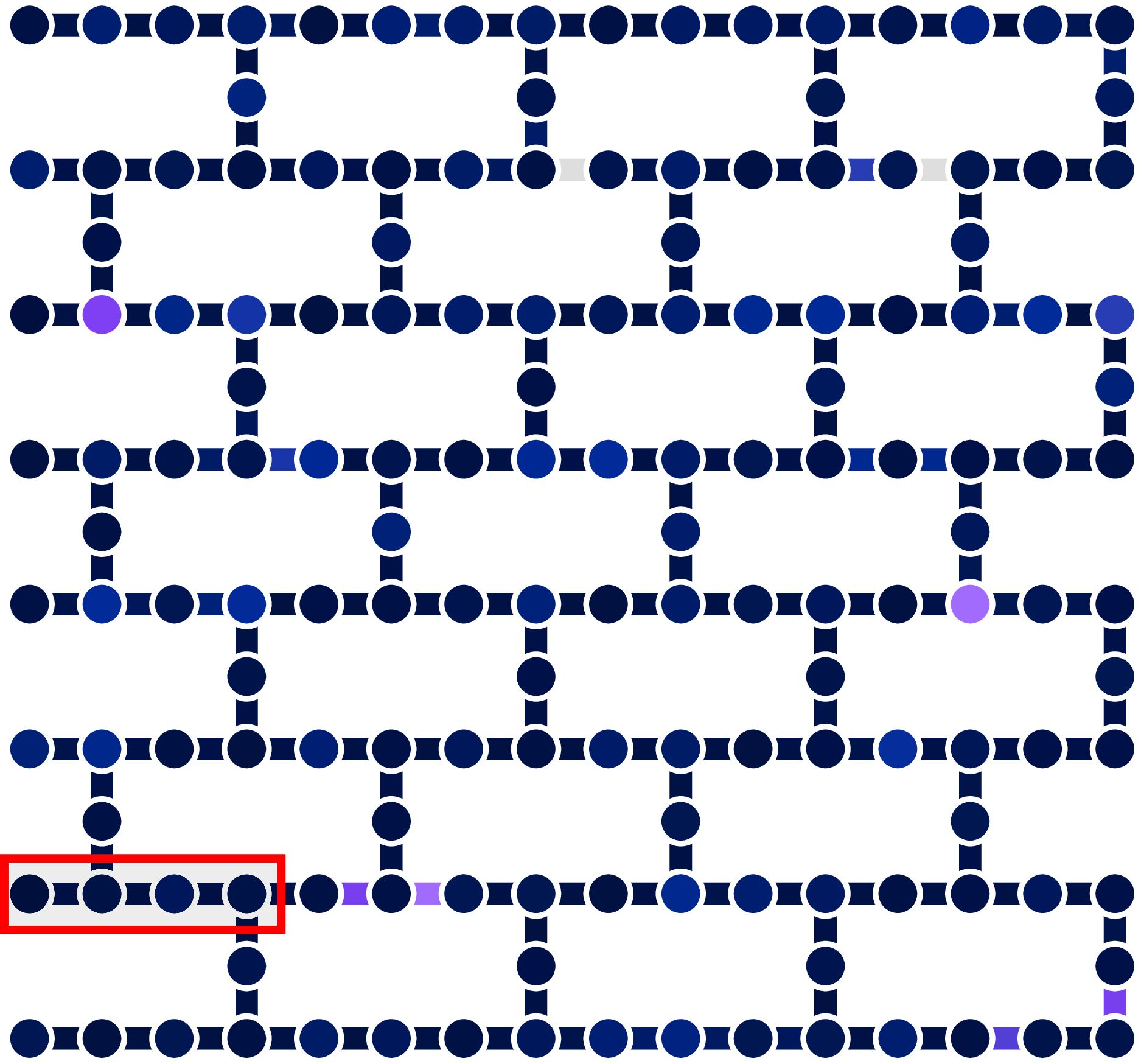}
    \caption{The IBM Fez QPU's qubit layout (vertices) and connectivity (edges) \cite{IBMQ2024}. This QPU is of the Heron family, comprised of $156$ qubits arranged in a heavy-hex lattice with cells of $12$ qubits. The entangling gates are $\operatorname{CZ}$ gates. The used qubits, numbers $120$ to $123$, were manually selected based on their readout and $\operatorname{CZ}$ errors at the time of the VQE execution.}
    \label{fig:ibm_fez}
\end{figure}

\begin{table}[t!]
    \centering
    \begin{tabular}{llcc}
        \cline{2-4} \\ [-2ex]
               & \multicolumn{1}{c}{$E$}     & $\Delta E$ & Iteration  \\ \hline  \\ [-1.5ex]
        $\text{min}(E_\text{QPU})$           & $-15.45925(651)$      & $0.10164$              & $167$ \\  [0.5ex]
        $\text{min}(E_\text{SVS})$           & $-15.55901$           & $0.00188$              & $139$ \\  [0.5ex] \hline \\ [-1.5ex]
        $\langle E_\text{QPU} \rangle_{\text{last}\ 10\%}$  & $-15.44416(879)$  & $0.11673$  & $163 - 180$   \\  [0.5ex]
        $\langle E_\text{SVS} \rangle_{\text{last}\ 10\%}$  & $-15.55824(26)$   & $0.00265$  & $163 - 180$ \\  [0.5ex] \hline
    \end{tabular}
    \caption{
        Summary of the QPU experiment's results. We report the minimum energies on QPU and corresponding SVS evaluation as well as their iteration numbers. We also show the averages over the last 10\% of iterations. All energies are given in Ha, and $\Delta E = |E - E^{\rm target}|$. The standard deviations given in the upper and lower halves of the table result, respectively, from the energy measurements and the averaging over the last 10\% of iterations.
    }
\end{table}

\begin{table}[t!]
    \centering
    \setlength{\tabcolsep}{10pt} 
    \begin{tabular}{llc}
        \cline{1-3} \\ [-2ex]
        Extrapolation   & \multicolumn{1}{c}{$E$ (Ha)}  & $\Delta E$ (Ha) \\ \hline  \\ [-1.5ex]
        $E_{\rm raw}$           & $-15.49705$      & $0.06196$ \\  [0.5ex]
        $E_{\rm lin}$           & $-15.69523$      & $0.13622$ \\  [0.5ex]
        $E_{\rm quad}$          & $-15.58634$      & $0.02733$ \\  [0.5ex]
        $E_{\rm exp}$           & $-15.60108$      & $0.04207$ \\  [0.5ex] \hline \\ [-1.5ex]
    \end{tabular}
    \caption{
        The ZNE-mitigated results using the parameters at iteration $139$. We show the results from three extrapolation methods: linear, quadratic, and exponential fittings. $\Delta E = |E - E_{\rm SVS}(\vtheta_{k=139})|$, with $E_{\rm SVS}(\vtheta_{k=139}) = -15.55901$ Ha.
    }
    \label{tab:zne-fez}
\end{table}

We ran a VQE using the \emph{HEA} on the IBM Fez QPU, described in Fig.\ref{fig:ibm_fez}, for $180$ iterations using the same setup that was described in previous subsections. The total computation time, including the SPSA calibration, classical pre-, post-processing and optimization, communication, and quantum computations, was 5h 30m 39s. The \emph{quantum time}, defined as the amount of time a QPU spends on performing a quantum computation task \cite{IBMQ2024}, totaled 1h 47m 02s.
Fig.\ref{fig:qpu_svs_energies} shows the results of the VQE run on IBM Fez. The minimum energy that was measured on the QPU was $E^{\rm min}_{\rm QPU} = -15.45925(651)$ Ha, at iteration $167$, which when evaluated on SVS gives $E_{\rm SVS}(\vtheta_{k=167}) = -15.55824$. However, when we evaluate each iteration's optimized parameters on SVS (without shot noise), we find that the best parameters are the ones produced at iteration $139$, with an SVS-evaluated energy of $E_{\rm SVS}(\vtheta_{k=139}) = -15.55901$ Ha. For reference, these parameters gave on QPU an energy of $E_{\rm QPU}(\vtheta_{k=139}) = -15.45790(650)$ Ha. The standard deviations given for the QPU energies are computed over the $4096$ shots of the energy measurements.
Finally, averaging over the last 10\% of iterations (18 in this case) for the QPU-estimated and SVS-estimated energies gives $\langle E_\text{QPU}(\vtheta_\text{QPU}) \rangle = -15.44416(879)$ Ha and $\langle E_\text{SVS}(\vtheta_\text{QPU}) \rangle = -15.55824(26)$ Ha respectively, where the standard deviations result from averaging over the 18 last energy values.

\subsubsection{Error Mitigation}
\label{subsubseq:error-mitigation}
When running a VQE on actual QPUs or noisy simulators, the raw energy obtained may be far from the ideal result due to the cumulative effects of errors. However, error mitigation (EM) techniques, such as zero-noise extrapolation (ZNE) \cite{PhysRevX.7.021050,9259940}, readout/measurement error mitigation \cite{PhysRevA.103.042605},  Clifford data regression \cite{czarnik2021error}, Pauli Twirling \cite{wallman2016}, or probabilistic error cancellation, can significantly improve the accuracy of the results by reducing the impact of noise on the final outcome.
To mitigate the effects of noise in the VQE's results of this study, we applied the ZNE error mitigation technique after the VQE. In the ZNE technique, the noise in quantum computations is artificially amplified, and the results are extrapolated back to the zero-noise limit to estimate ideal noiseless outcomes. As error mitigation is not the focus of this work, the reader may refer to \citet{9259940} for further details on the this method.\\

For our error mitigation step, we used the parameters corresponding to iteration $139$—the iteration with the best SVS-evaluated energy, $E_{\rm SVS}(\vtheta_{k=139}) = -15.55901$. ZNE was carried out on the same QPU as the VQE, IBM Fez, using $40,000$ shots per circuit, with integer noise-scaling factors (folds) $1$, $3$, and $5$. The raw energy without any mitigation, $E_{\rm raw}$, or fold $1$, was measured to be $-15.49705$ Ha, with an absolute error of $\Delta E = 61.96$ mHa with respect to the above target energy.
Table \ref{tab:zne-fez} summarizes the results of three extrapolations using a linear, quadratic, and exponential fitting functions. We find for our case that the quadratic extrapolation achieved the best accuracy with $E_{\rm quad} = -15.58634$ Ha and $\Delta E = 27.33$ mHa.
Note that for the extrapolation procedures, we use the average measured energy values only without taking into account their standard deviations, and we thus report the ZNE results without standard deviations. Another point to take into consideration is that due to possible changes in the QPU's noise characteristics between the VQE and ZNE experiments, the parameters, $\vtheta_{k=139}$, may yield different values in the VQE and ZNE measurements. Consequently, $E_\text{QPU}(\vtheta_{k=139})$ with fold $1$ was measured again at the same time as the other folds, so that all folds are affected by the same noise and device characteristics.

\section{Discussion}
\label{sec:discussion}


In the context of the VQE, \emph{ideal simulations} refer to the computation of the energy using a noiseless quantum circuit, typically carried out via a noiseless state-vector simulator (SVS). This method provides the theoretical ground state energy that would be obtained if all quantum gates and measurements were executed perfectly without any decoherence, gate errors, or readout errors. We do, however, simulate the fluctuations in quantum measurements, known as shot noise, in the SVS VQEs. We remind that the target energy for our molecular problem is $E^{\rm target} = -15.56089$ Ha, and we give here, for reference, the Hartree-Fock energy as $E^{\rm HF} = -15.56033$ Ha.\\
Noiseless simulations clearly demonstrate the reliability of the \emph{Unitary Coupled-Cluster Single and Double} excitations (\emph{UCCSD}) anstaz, with the majority of converging VQE instances reaching chemical accuracy ($1.6$ mHa from the target energy) after fewer iterations compared to the \emph{Hardware-Efficient Ansatz} (\emph{HEA}) as shown on Fig.\ref{fig:all_in_one_with_convergence}. 
Moreover, it is noteworthy that \emph{UCCSD} provides an order of magnitude better average energy value ($\Delta E = 0.05$ mHa) compared to the \emph{HEA} ($\Delta E = 0.34$ mHa), which highlights the efficiency of the chemically inspired UCC theory-based ansatz in the absence of noise.
Additionally, in the absence of noise, both ansätze yield energy estimates within chemical accuracy of the target and below the Hartree-Fock (HF) energy, showcasing the effectiveness of the VQE under ideal, noiseless conditions.\\

However, real-world quantum computers introduce noise into the computation due to imperfections in gate operations, decoherence, environment-induced noise, and measurement errors. In noisy simulations, the energy measured is generally higher than the SVS energy, reflecting these additional imperfections. Therefore, comparing the ideal $E_{\text{SVS}}$ with the energies obtained from noisy runs provides insight into optimization process under noise and the usefulness or limitations of current hardware. 
In this study we compared the computational accuracy at which the ground state energy of the \ce{BeH2} molecule can be estimated on three different quantum computer noise models for: IBM Strasbourg, Torino, and Fez. Each of these exhibiting distinct error rates.
The effect of noise pushes the energy values above chemical accuracy by two and four orders of magnitude for the \emph{HEA} and the \emph{UCCSD}, respectively, when compared to the ideal device simulations. 
The difference becomes evident when comparing the performance of the \emph{HEA} to the \emph{UCCSD} ansatz. Errors are an order of magnitude higher for the \emph{UCCSD} ansatz, independent of the noise model. This discrepancy is largely attributed to the significant difference in circuit depths between the two ansätze (see Table \ref{table:ansatz}), highlighting the better noise-resilience of the \emph{HEA} and emphasizes \emph{UCCSD}'s sensitivity to hardware noise. Moreover, the absolute error across the three devices is of the order of $10^{-2}$ Ha for \emph{HEA} but rises to the order of $10^{-1}$ Ha for \emph{UCCSD}. This proves the greater robustness of \emph{HEA} to hardware noise. It is also noteworthy that \emph{UCCSD} exhibits larger measurement fluctuations in noisy simulations.
%
Interestingly, the average energy evaluated on SVS over the set of the last 10 \% of parameters, $\{\vtheta_{\text{best}}\}_{\rm last 10\%}$, for best performing noisy VQEs serves as a reference for what the variational ansatz could achieve under ideal conditions.
When evaluated on SVS, all the results of the \emph{HEA}-based VQE are within chemical accuracy from the energy target. Meanwhile, the \emph{UCCSD} energy values remain beyond chemical accuracy. However, the error with respect to the target energy was reduced by one order of magnitude for IBM Strasbourg and two orders of magnitude for Torino and Fez. 
This shows the different effects of noise on the quality of the optimized parameters on one hand, and on the accuracy of the evaluated energy from the obtained VQE parameters on the other. 
A more in-depth analysis of VQE's performance across different noise levels on the three noisy simulators we used is beyond the scope of this paper and will be addressed in future work.

In the light of the previous simulations, the results of the VQE implementation on IBM Fez, shown in Fig.\ref{fig:qpu_svs_energies}, are particularly interesting. The minimum energy that was measured on the QPU was $E_{\rm QPU}^{\rm min}= -15.45925(651)$ Ha, corresponding to the parameters at iteration $167$, $\vtheta_{k=167}$. When evaluated on SVS, these same parameters result in an evaluated energy $E_{\rm SVS}(\vtheta_{k=167}) = -15.55847$ Ha, higher than the target energy value by $2.24$ mHa. However, evaluating the parameters $\vtheta_{\rm QPU}$ for all iterations on SVS shows that a better parameter vector, $\vtheta_{k=139}$, has an energy $E_{\rm SVS}(\vtheta_{k=139}) = -15.55901$ Ha, a mere $1.88$ mHa above the target energy. This finding indicates that we may optimize parameters well on QPU, despite misestimating their energies.
Moreover, the average energy over the last 10\% of iterations (18 in this case) for the  SVS-evaluated energies is $\langle E_\text{SVS}(\vtheta_\text{QPU}) \rangle_{\text{last}\ 10\%} = -15.55824(26)$ Ha, which is within the same range of 2$\times$ chemical accuracy from the target energy ($[-15.56089, -15.55769]$ Ha).
These SVS-evaluated energies draw a better picture of the quality of the solution produced by the VQE compared to the QPU-estimated energy value, and show that the VQE did converge to a good solution despite quantum noise and the larger error in the estimation of energy values by the QPU.
The average QPU-estimated energy over the last 10 \% of iterations was $\langle E_\text{QPU}(\vtheta_\text{QPU}) \rangle_{\text{last}\ 10\%} = -15.44416(879)$ Ha, again significantly higher than what the parameter vectors would give on SVS for the same iterations. This average also displays a larger standard deviation as noise amplifies the fluctuations in the energy estimation.

After applying error mitigation, we obtain a corrected energy value which serves as a more accurate approximation of the true ground state energy in the presence of noise. The {mitigated energy} should be regarded as one of the key results in assessing the success of the VQE experiment, as it reflects both the experimental realities of running quantum circuits on noisy hardware and the effectiveness of the error mitigation strategies employed.

In our work, we demonstrated the use of zero-noise extrapolation (ZNE) on real quantum hardware to mitigate the measured QPU energy values. The error mitigation results presented in Table. \ref{tab:zne-fez} show various degrees of improvements to the QPU-measured energy. The quadratic and exponential extrapolations improved upon the unmitigated QPU energy yielding respectively absolute errors $\Delta E_{\rm quad} = 27.33$ mHa, and $\Delta E_{\rm exp} = 42.07$ mHa. The linear extrapolation however produced a significantly worse error, $\Delta E_{\rm lin} = 136.22$ mHa. These results showcase the ability of methods such as ZNE to correct to a certain degree for the effect of noise on the quality of measured energies on noisy QPUs. This improvement is however not guaranteed. A poor choice of extrapolation methods, as was the case in the linear extrapolation for this specific case, will produce poor results. This, in particular, is one of the weaknesses of ZNE. Other techniques such as Clifford data regression aim to address these shortcomings, with challenges of their own \cite{czarnik2021error}.

\begin{figure}[t!]
    \centering
    \includegraphics[width=1\linewidth]{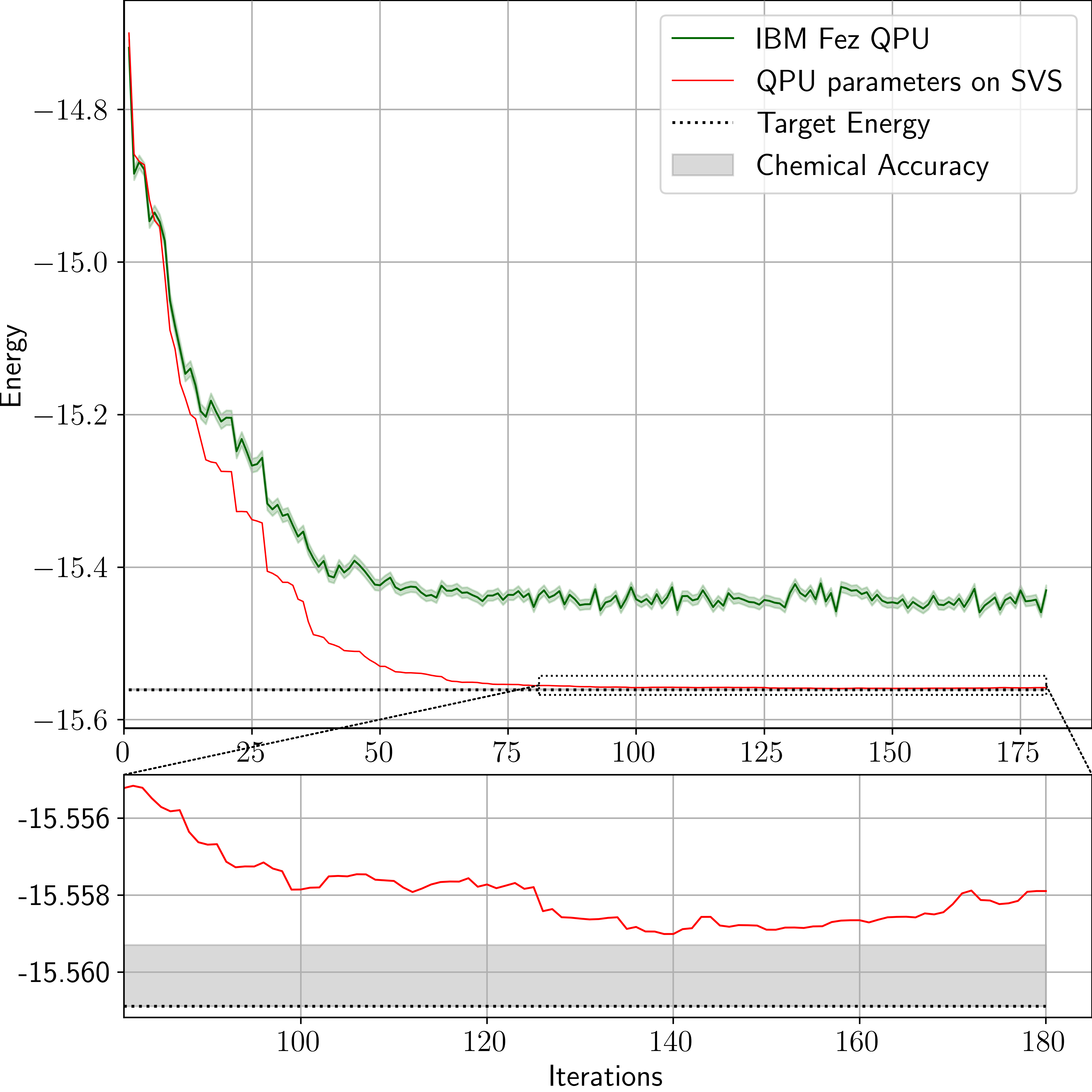}
    \caption{
        Results of the VQE run on IBM Fez (green). At each iteration, the optimized parameters are also evaluated on an SVS. The shaded area around the QPU energy graph is the standard deviation of the QPU measurements at 4096 shots. The corresponding SVS-evaluated energy graph (red) for the same optimized parameters is also shown.
    }
    \label{fig:qpu_svs_energies}
\end{figure}

\begin{figure}[t!]
    \centering
    \includegraphics[width=1\linewidth]{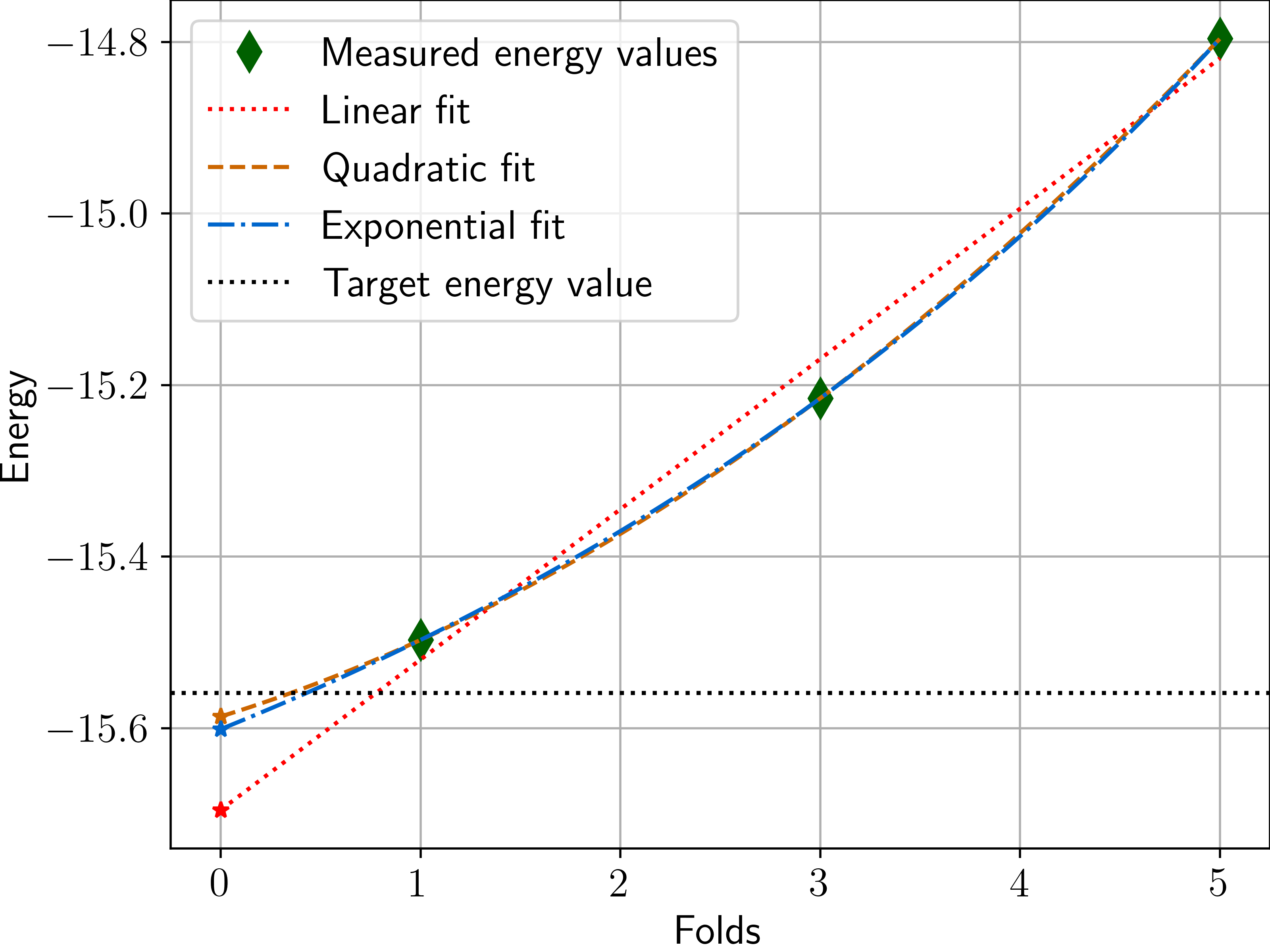}
    \caption{
        Zero-noise extrapolation results for the parameters of iteration $139$. We show three fittings: linear, quadratic, and exponential, in addition to the zero-noise extrapolations (stars) at $x=0$.
    }
    \label{fig:qpu_zne}
\end{figure}

\section{Conclusion}

We have presented a comprehensive and practical guide that balances the depth of review articles with the conciseness of shorter communications and guides for the implementation of the Variational Quantum Eigensolver (VQE) to estimate the ground state energy of the \ce{BeH2} molecule. 
Our detailed study addresses several aspects of the end-to-end VQE implementation that we found insufficiently documented in existing literature. These include the construction of the electronic Hamiltonian, the mapping of the Hamiltonian to qubit operators via second quantization, and the computation of one- and two-electron integrals with detailed calculations exemplified through the \ce{H2} molecule. We also elaborated on the mathematical framework for the Unitary Coupled Cluster with Single and Double excitations (\emph{UCCSD}) and provided an updated methodology for implementing the VQE using the latest version of \texttt{Qiskit (1.2)} employing the Simultaneous Perturbation Stochastic Approximation (SPSA) as classical optimizer. 
Our algorithm, run on both ideal and noisy simulators, as well as on a real quantum device, successfully converges toward the target energy estimated from classical calculations within a reasonable number of iterations without requiring error mitigation during the VQE implementation. This work aims to provide a theoretical background and to provide essential tools for the simulation of larger molecules using the VQE.
To demonstrate the effectiveness of the VQE on currently available quantum hardware, we performed energy calculations using noiseless simulators and noisy simulators based on the characteristics of three IBM quantum devices, each with distinct error rates: IBM Strasbourg, Torino, and Fez. Additionally, we carried out computation on the IBM Fez quantum computer, the most advanced device available to us with the lowest noise level, and consisting of 156 qubits.

Our study presents a comparative analysis of two conceptually different ansätze: the chemically inspired \emph{UCCSD} and a hardware-efficient ansatz (\emph{HEA}). While \emph{UCCSD} achieves a higher accuracy on the state-vector simulator (SVS), it is significantly more sensitive to noise, making it less suitable for current NISQ devices. In contrast, the \emph{HEA} exhibits promising performance across all platforms—SVS, noisy simulators, and actual quantum hardware. Notably, \emph{HEA} effectively optimizes parameters, achieving energy estimates within chemical accuracy relative to the exact solution at the level of the employed theory on SVS. Additionally, across all noisy simulations, \emph{HEA} remains robust, and on quantum processing units (QPUs), it produces optimized ground states corresponding to an exact energy estimate only $1.88$ mHa above the target energy, even without the application of error mitigation (EM) techniques.

Indeed, while error mitigation techniques have proven highly effective in enhancing the accuracy of the VQE on current quantum computers, they often come at the cost of significantly increased resource demands. It remains uncertain whether this added resource requirement will be a manageable trade-off or a critical limitation as the VQE is scaled to larger, more complex applications. Our findings demonstrate that achieving ground state energy within chemical accuracy, compared to the exact solution at the chosen level of theory, is feasible without needing error mitigation during the VQE convergence. Applying EM as a post-processing step can significantly reduce the computational resources required.

Furthermore, state-vector energy estimations using the quantumly optimized parameters confirm that current quantum devices are effective in optimizing circuit parameters despite their tendency to misestimate the actual values of simulated energies. Similar results were reported by \citet{sorourifar2024, Sorourifar2024-0} using Bayesian optimization, while we observe this trend with SPSA in our study. The higher accuracy in estimating the energy landscape features over energies themselves thus appears independent of the optimizer used. We plan to explore this observation further in future works.\\

\section*{Acknowledgments}
This document has been produced with the financial assistance of the European Union (Grant no. DCI-PANAF/2020/420-028), through the African Research Initiative for Scientific Excellence (ARISE), pilot programme. ARISE is implemented by the African Academy of Sciences with support from the European Commission and the African Union Commission. The contents of this document are the sole responsibility of the author(s) and can under no circumstances be regarded as reflecting the position of the European Union, the African Academy of Sciences, and the African Union Commission.
The authors thank the Algerian Ministry of Higher Education and Scientific Research and DGRSDT for financial support. 
We acknowledge the support of the Quantum Collaborative for their support and access to IBM Quantum Resources. 
D.B.N. is supported by the startup grant of the Davidson School of Chemical Engineering at Purdue University.

\vfill{}
\bibliography{bib.bib}

\appendices
\section{Molecular orbitals construction}
\label{app:LCAO+basis-set}

To represent the spatial distribution of electrons in molecules, we need to choose an orthonormal basis set for the molecular orbital functions $\{\xi_p(\mathbf{r})\vert p=1\cdots M\}$. In computational chemistry, it is convenient to construct those molecular orbitals based on our knowledge of atomic orbitals $\{\phi_\alpha(\mathbf{r})\vert  \alpha=1\cdots M\}$. In the Linear Combination of Atomic Orbitals (LCAO) method, molecular orbitals $\xi_p$ are expressed as a linear combination of atomic orbitals $\phi_\alpha$. Each atomic orbital has a real coefficient $c_{p\alpha}$ that represents its contribution to the molecular orbital. Each molecular orbital is then expressed as
\begin{equation}
	\xi_p(\mathbf{r}) = \sum_\alpha c_{p\alpha} \phi_\alpha(\mathbf{r}).
    \label{eq:lcao-mo}
\end{equation}
However, since it is hard to compute the electron integrals, especially the two-electron integrals, with Hydrogen-like atomic orbitals that have a Slater determinant form,
it is more convenient to approximate these orbitals with a linear a combination of normalized primitive Gaussian functions \cite{hehre1969} which take the form
\begin{equation}
    \sigma_c(x,y,z) = N_c (x-R_x)^i (y-R_y)^j (z-R_z)^k \exp^{-\alpha_c (\mathbf{r}-\mathbf{R})^2},
    \label{eq:primitive-gaussian}
\end{equation}
where $i,j,k$ are non-negative integers and the orbital number $l=i+j+k$  specifies the shell type of the spherical part of the wave-function. The normalization factor $N_c$ is given by \cite{cramer2004, hehre1969}
\begin{equation}
    N_c=\left(\frac{2\alpha_c}{\pi}\right)^{3/4} \left(\frac{(8\alpha_c)^{i+j+k} i!j!k!}{(2i)!(2j)!(2k)!}\right)^{1/2}.
    \label{eq:norm-factor}
\end{equation}
The atomic orbitals are thus approximated as follows:
\begin{align}
    \phi_\alpha(\mathbf{r}) = \sum_{c} d_{\alpha c} \sigma_c(\mathbf{r}).
    \label{eq:ao-approx}
\end{align}
While the coefficients $d_{ac}$ in Eq.\eqref{eq:ao-approx} and the exponents $\alpha_c$ in Eq.\eqref{eq:primitive-gaussian} and Eq.\eqref{eq:norm-factor} are determined to approximate the conventional atomic orbitals and preserve the normalization, the $c_{\mu a}$ coefficients in eq. \eqref{eq:lcao-mo} are computed with the Self-Consistent Field (SCF) method where the mean field energy of the molecule is minimized to get the  Hartree-Fock state and the coefficients of the orbitals \cite{lewars2010, cramer2004}.
Basically, the orbital wave functions are written as a linear combination of the basis set as follows:
\begin{align}
    \label{eq:LCAO-STOnG}
    \xi_p(\mathbf{r}) = \sum_{\alpha c} c_{p\alpha} d_{\alpha c} \sigma_c(\mathbf{r}).
\end{align}
However, since electrons are spin $1/2$ particles, their wave function should include a spin factor $\alpha(\mathbf{r})$ or $\beta(\mathbf{r})$ for a spin up or down, respectively. Therefore, the final form of the spin molecular orbitals is:
\begin{align}
    \psi_p(\mathbf{x}) = \psi_p(\mathbf{r}, s) = 
    \left\{
    \begin{array}{c}
         \xi_p(\mathbf{r}) \alpha(\mathbf{r})_{s=\uparrow}\\
         \xi_p(\mathbf{r}) \beta(\mathbf{r})_{s=\downarrow}
    \end{array}
    \right.
\end{align}
Taking into consideration the fermionic statistics of electrons, the Hartree-Fock method gives the wave function of the electronic ground state $\Psi(\mathbf{x_1, \cdots, x_N})$ as the Slater determinant of the spin molecular orbitals $\{\psi_p(\mathbf{x}_i)\vert p=1,\cdots, 2M; i=1,\cdots, N\}$ for $N$ electrons and $M$ molecular orbitals such as $2M>N$.\\

Choosing a basis set is crucial to get an accurate estimation of the ground state energy \cite{dobrautz2024}. The most accurate results can be achieved by considering all combinations of interactions between electrons in different molecular orbitals, all possible Slater determinants, and large basis set expansions \cite{Szabo1996,lewars2010}. However, this leads to more Hamiltonian terms and requires vast computational resources.
There exists a diversity of basis sets in quantum chemistry literature, each offering advantages and disadvantages that depend on the nature of the molecule to be studied. We cite three examples here which are suitable for small, medium, and large molecules :

\begin{itemize}
\item Minimal-Basis Sets: the STO-3G basis set \cite{hehre1969, Szabo1996-basis-sets, lewars2010-basis-sets} is one of the simplest options that is widely used for small molecules. It's a linear combination of three Gaussian functions of the form $d \cdot \exp(-\alpha\mathbf{r}^2)$ that produces Slater-type orbitals. The orbitals are distributed over the conventional shells. For the first row of atoms, we only have the $1s$ orbital. For the second row atoms, we have the $1s$ for the first shell and the $2s$ \& $2p$ orbitals for the second shell. In general, the orbitals that share the same shell are given the same Gaussian exponents. For more accuracy, it is possible to use an STO-$n$G basis set, with a $n>3$, which uses a linear combination of $n$ Gaussian functions for each orbital.
The minimal-basis sets give reliable results within short computational times, still, the accuracy not enough for molecules with more electrons and atoms.

\item Small and Medium Basis Sets: the 3-21G, 3-21G${}^{(*)}$, and 6-31G basis sets, or X-YZG \cite{hehre1969, lewars2010-basis-sets}, are used for medium-sized molecules. These basis sets utilize two sets of functions, one for the core orbitals (X Gaussian functions), while each valence orbital is split into an inner one with Y Gaussian functions and an outer orbital with Z Gaussian functions (one in the above examples). 
Such medium basis sets provide more accuracy than STO-3G but will require more computational power. The previous two basis set types are introduced by Pople and his group \cite{hehre1969}.

\item Correlation-Consistent Basis Sets: the cc-sets are larger sets that are suitable for accurate chemical computations; they are called cc$pVXZ$ and first introduced by Dunning \cite{lewars2010-basis-sets, Dunning1989}, where $p$ stands for polarization functions, $V$ for valence, $X$ for the number of shells the valence functions are split into, and $Z$ for zeta. For instance, the cc-pVTZ means correlation-consistent polarized valence triply-split zeta. These sets are more computationally demanding than either STO-\textit{n}G and \textit{X}-\textit{YZ}G but are significantly more accurate.
\end{itemize}

\section{Building the Second Quantized Form}
\label{app:second-quantized-hamiltonian}
Since electrons are indistinguishable, we do not care about which electron occupies which orbital. The second quantized states involve only information about the occupied orbitals. We start from the vacuum state $\ket{\text{vac}}$ where all orbitals are not occupied, and there is no electron. Then, we start filling orbitals by creating electrons using the creation operators $a_p^\dagger$ for each spin molecular orbital $\psi_p$. Namely, $a_p^\dagger\ket{\text{vac}} = \ket{\psi_p}$. However, the creation of two electrons in the system implies an anti-symmetric state given by the Slater determinant:
\begin{align}
a_q^\dagger a_p^\dagger\ket{\text{vac}}
&= \frac{1}{\sqrt{2}}\left( \ket{\psi_p}\ket{\psi_q} - \ket{\psi_q}\ket{\psi_p} \right).
\end{align}
The state $\ket{\psi_q}\ket{\psi_p}$ means simply that the first electron is in the state $\ket{\psi_p}$ while the second electron in the state $\ket{\psi_q}$. Hence, the creation operators should obey the following algebra:
\begin{align}
    \{a_p^\dagger, a_q^\dagger\} &= 0,
\end{align}
and the same for the annihilation operators $a_p$ and  $a_q$.  Thus, the fermionic statistics are obeyed. The algebra of ladder operators is complete with :
\begin{align}
    \{ a_p^\dagger, a_q\} &= \delta_{pq},
\end{align}
that accounts for Pauli's exclusion principle. Generally, we define the Fock state as
\begin{align}
    \vert n_0 n_1 \cdots n_{k}\rangle 
    &= (a_0^{\dagger})^{n_0} (a_1^{\dagger})^{n_1} \cdots (a_k^{\dagger})^{n_k} \ket{\text{vac}},
\end{align} 
such that
\begin{align}
    n_i = \left\{
    \begin{array}{ll}
        1 & \text{if the $i$'th spin molecular orbital is occupied,} \\
        0 & \text{Otherwise}
    \end{array}\right.
\end{align}
The state $\vert n_0 n_1 \cdots n_{k}\rangle$ is a compact representation of a Slater determinant of all the occupied modes.

\subsection{One-Electron Terms}
The one-electron terms in the Hamiltonian are the kinetic term and the nucleus-electron Coulomb interaction term, that take the form:
\begin{align}
    \hat{F}      &= \sum_{i=1}^N \hat{f}(i).
\end{align}
such that $\hat{f}(i)$ is a function of the $i$'th electron's momentum and position operators. For each electron, the spin molecular orbitals form an orthonormal basis:
\begin{align}
    \begin{array}{cc}
    \sum_{p} \ket{\psi_p}_i \prescript{}{i}{\bra{\psi_p}} = 1  & \forall \text{ electron } i.
    \end{array}
\end{align}
Therefore, we can write the operator $\hat{f}(i)$ as
\begin{align}
    \hat{f}(i)
    &= \sum_{pq} \prescript{}{i}{\bra{\psi_p}} \hat{f}(i) \ket{\psi_q}_i \ket{\psi_p}_i  \prescript{}{i}{\bra{\psi_q}}\\
    &= \sum_{pq} f_{pq} \ket{\psi_p}_i  \prescript{}{i}{\bra{\psi_q}},
\end{align}
since $\prescript{}{i}{\bra{\psi_p}} \hat{f}(i) \ket{\psi_q}_i = f_{pq}$ what ever $i$ is, the one-electron operators will be:
\begin{align}
    \hat{F}      
    &= \sum_{i=1}^N \sum_{pq} {f}_{pq} \ket{\psi_p}_i  \prescript{}{i}{\bra{\psi_q}}\\
    &= \sum_{pq} {f}_{pq} \sum_{i=1}^N \ket{\psi_p}_i  \prescript{}{i}{\bra{\psi_q}}.
\end{align}
It is possible to prove by involving an accurate correspondence between Fock and orbital states \cite{Tannoudji2017-tome3} that:
\begin{align}
    \sum_{i=1}^N \ket{\psi_p}_i  \prescript{}{i}{\bra{\psi_q}} = a_p^\dagger a_q.
\end{align}
Therefore,
\begin{align}
    \hat{F} = \sum_{pq} f_{pq} a_p^\dagger a_q.
\end{align}
And since the operator $\hat{f}$ terms are functions of momentum and position operators, 
\begin{align}
    f_{pq} &= \bra{\psi_p} \hat{f} \ket{\psi_q}\\
    &= \int \psi_p^{*}(\mathbf{r}) \hat{f} \psi_q(\mathbf{r}) d\mathbf{r}.
\end{align}

\subsection{Two-Electron Terms}
The general form of a two-body operator, such as the Coulomb interaction between two electrons, can be written as
\begin{align}
    \hat{G} &= \sum_{\substack{i,j=1\\i > j}}^N \hat{g}(i, j).
\end{align}
Knowing that any two-body operator can be written as an expansion of the product of two one-body operators:
\begin{align}
    \hat{G}
    &= \sum_{\substack{i,j=1\\i > j}}^N \sum_k c_k \hat{f}^{\alpha_k}(i) \hat{h}^{\beta_k}(j)\\
    &= \sum_k c_k \sum_{\substack{i,j=1\\i > j}}^N \hat{f}^{\alpha_k}(i) \hat{h}^{\beta_k}(j).
\end{align}
Using the results of one-electron operator and from the algebra of ladder operators:
\begin{align}
    a_p^\dagger a_s a_q^\dagger a_r = a_p^\dagger a_q^\dagger a_r a_s + \delta_{qr} a_p^\dagger a_s,
\end{align}
it is possible to show that
\begin{align}
    \hat{G} = \sum_{pqrs} g_{pqsr} a_p^\dagger a_q^\dagger a_s a_r
\end{align}
where
\begin{align}
\begin{array}{ll}
g_{pqsr} = \prescript{}{i}{\bra{\psi_p}} \prescript{}{j}{\bra{\psi_q}} \hat{g}(i,j) \ket{\psi_s}_j \ket{\psi_r}_i
 & \forall i,j.
\end{array}
\end{align}
Finally, we re-write the above as
\begin{align}
g_{pqrs} = \int \int \psi_p^{*}(\mathbf{r_1}) \psi_q^{*}(\mathbf{r_2}) \hat{g}(1,2) \psi_s(\mathbf{r_2}) \psi_r(\mathbf{r_1}) d\mathbf{r_1}d\mathbf{r_2}.
\end{align}

\onecolumngrid
\section{Computing the one- and two-electron integrals for the \ce{H2} molecule}
\label{app:H2-integrals}
In this appendix, we will compute, as an example, the one- and two-electron integrals for the H2 molecule with an interatomic distance of $0.74 \AA$ using the STO-3G basis set.

\subsection{Molecular geometry of the H\tsb{2} molecule}
\quad 
\begin{figure}[h]
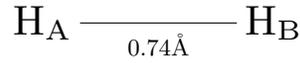

\centering
	\vspace{1cm}
	\setchemfig{bond offset=3pt, atom sep=50pt, atom style={scale=1.75}}
	\chemfig{@{ha}H\tsb{A}-@{hb}H\tsb{B}}
	\namebond{ha}{hb}{below}{$0.74 \AA$}
	\vspace{0.5cm}
	\caption{The molecular geometry of H\tsb{2}.}
	\label{fig:H2_geometry}
\end{figure}
\quad The molecular geometry of the H\tsb{2} molecule is rather simple, as illustrated in FIG. \ref{fig:H2_geometry}, it is only two hydrogen atoms separated by a $0.74 \AA$ bond distance at the equilibrium ground state; the first quantized electronic Hamiltonian for this molecule is given in the following form:

\begin{equation}    
H_{el} = - \frac{\nabla_{1}^2}{2} - \frac{\nabla_{2}^2}{2} -  \frac{1}{|\mathbf{R}_a - \mathbf{r}_1|} - \frac{1}{|\mathbf{R}_b - \mathbf{r}_1|} - \frac{1}{|\mathbf{R}_a - \mathbf{r}_2|} - \frac{1}{|\mathbf{R}_b - \mathbf{r}_2|}  +  \frac{1}{|\mathbf{r}_1 - \mathbf{r}_{2}|}.
\end{equation}
It is worth it to note that the Hamiltonian is in the atomic units. The one-electron integrals $h_{pq}$ are defined as in Eq. \eqref{eq:h_pq}. Practically, since the spin states are orthonormal, we write these integrals in the molecular orbitals basis $\{\xi_p(\boldsymbol{r})\}$ in two parts as $h_{pq} = T_{pq} + V_{pq}$ such that:
\begin{align}
T_{pq} &= -\frac{1}{2} \int \xi_p(\mathbf{r}) \nabla^2 \xi_q(\mathbf{r}) \, d\mathbf{r},\\
V_{pq} &= \sum_{c\in \text{nuclei}} \int \xi_p(\mathbf{r}) \frac{-1}{|\mathbf{r} - \mathbf{R}_c|} \xi_q (\mathbf{r}) \, d\mathbf{r}, 
\end{align}
where $T_{pq}$ and $V_{pq}$ represent the contribution of the $p$ and $q$ molecular orbitals to the kinetic energy and the nuclear attraction energy, respectively. The two-electron integrals in Eq. \eqref{eq:h_pqrs} can be written in the molecular basis since the two spin states are orthonormal to each other as:
\begin{equation}
\label{eq:h_pqrs-H2}
h_{pqrs} = \int \int \xi_p(\mathbf{r}_1) \xi_q(\mathbf{r}_2) \frac{1}{|\mathbf{r}_1 - \mathbf{r}_2|} \xi_r(\mathbf{r}_1) \xi_s(\mathbf{r}_2) d\mathbf{r}_1 d\mathbf{r}_2,
\end{equation}
where $\xi_p(r)$ are the molecular orbitals' functions described as a LCAO as shown in Eq. \eqref{eq:lcao-mo}. Working in the STO-3G basis, we define our atomic orbitals (AO) as a linear combination of three normalized Gaussian functions:

\begin{equation}
    \label{eq:H2-AO-STO-3G}
    \phi_\alpha(\mathbf{r}) = \sum_{c=1}^{3} d_{\alpha c} \sigma_c(\mathbf{r}-\mathbf{R}_\alpha),
\end{equation}
where the normalized Gaussian functions \eqref{eq:primitive-gaussian} are given in the case of $s$ orbitals ($l=0$) as:

\begin{equation}
    \sigma_c(\mathbf{r}-\mathbf{R}_\alpha) = 
    \left(\frac{2\alpha_c}{\pi}\right)^{3/4} \exp({-\alpha_c(|\mathbf{r}-\mathbf{R}_\alpha)|^2}).
\end{equation}

\subsection{Kinetic Energy Integral Computation Over Atomic Orbitals in the STO-3G basis}

The computation of kinetic energy integrals over primitive atomic orbitals in the STO-3G basis is performed using the following integral:

\begin{equation}
T_{pq} = -\frac{1}{2} \sum_{\alpha \beta=1}^2 c_{p\alpha} c_{q\beta} \int \phi_\alpha(\mathbf{r}) \nabla^2 \phi_\beta(\mathbf{r}) \, d\mathbf{r},
\end{equation}
where $\phi_a$ and $\phi_b$ are AO functions defined in Eqs. \eqref{eq:lcao-mo} and \eqref{eq:H2-AO-STO-3G}. In the STO-3G basis, the kinetic energy integral can be rewritten by developing the AO as a linear combination of Gaussians:
\begin{equation}\label{T_pq}
T_{pq} = \sum_{\alpha\beta=1}^{2} \sum_{ab=1}^{3} c_{p\alpha} c_{q\beta} d_{\alpha a} d_{\beta b} \langle a | -\frac{1}{2} \nabla^2 | b \rangle,
\end{equation}
where:
\begin{equation}
    \langle a | -\frac{1}{2} \nabla^2 | b \rangle = \int -\frac{1}{2} \sigma_a(\mathbf{r}-\mathbf{R}_a) \nabla^2 \sigma_b(\mathbf{r}-\mathbf{R}_b) \, d\mathbf{r}.
\end{equation}
The following computations of one- two-electron integrals over primitive Gaussian functions follow the methodology outlined in \citet{Szabo1996}.\\
Evaluating $\langle a | -\frac{1}{2} \nabla^2 | b \rangle $ is straight forward after letting $\nabla^2$ operate, we utilize the Gaussian product theorem to express the two Gaussians as one centered in $\mathbf{R}_p$:
\begin{equation}
    \sigma_a(\mathbf{r}-\mathbf{R}_a) \sigma_b(\mathbf{r}-\mathbf{R}_b) =  N_p \exp({-\alpha_p|\mathbf{r}-\mathbf{R}_p|^2}) \exp\left(-\frac{\alpha_a \alpha_b}{\alpha_a + \alpha_b} |\mathbf{R}_a - \mathbf{R}_b|^2\right),
\end{equation}
where the new exponent $\alpha_p=\alpha_a+\alpha_b$ and $N_p=N_aN_b$ and $\boldsymbol{R_p}$ takes the form:
\begin{equation}
\mathbf{R}_p = \frac{\alpha_a \mathbf{R}_a + \alpha_b \mathbf{R}_b}{\alpha_a + \alpha_b}.
\end{equation}
Simplifying and evaluating the integrals, $\langle a | -\frac{1}{2} \nabla^2 | b \rangle $ takes the form:
\begin{equation}
    \langle a | -\frac{1}{2} \nabla^2 | b \rangle =N_aN_b \frac{\alpha_a \alpha_b}{\alpha_a + \alpha_b}\left[3- \frac{2\alpha_a \alpha_b}{\alpha_a + \alpha_b} |\mathbf{R}_a - \mathbf{R}_b|^2 \right]\left[ \frac{\pi}{\alpha_a + \alpha_b}\right]^{\frac{3}{2}} \exp\left(-\frac{\alpha_a \alpha_b}{\alpha_a + \alpha_b} |\mathbf{R}_a - \mathbf{R}_b|^2\right).
\end{equation}
We can get $T_{pq}$ by summing over all contractions as shown in Eq. \eqref{T_pq}.

\subsection{Nuclear Attraction Integral Computation Over Atomic Orbitals in the STO-3G basis}

\label{sec:one-electron-nuclear-integral-H2}
The computation of the nuclear attraction integrals over primitive Gaussian functions can be derived similarly to the kinetic energy integral. The integral to be evaluated is:
\begin{equation}
V_{pq}= \sum_{c\in\text{nuclei}} \sum_{\alpha \beta=1}^2 c_{p\alpha} c_{q\beta} \int \phi_\alpha(\mathbf{r}) \frac{-1}{|\mathbf{r} - \mathbf{R}_c|} \phi_\beta(\mathbf{r}) \, d\mathbf{r}.
\end{equation}
We can develop the AO in the STO-3G basis to get:
\begin{equation}\label{V_pq}
V_{pq}= \sum_{c\in\text{nuclei}} \sum_{\alpha \beta=1}^2 c_{p\alpha} c_{q\beta} \sum_{ab=1}^{3} d_{pa} d_{qb}   \langle a | \frac{-1}{|\mathbf{r} - \mathbf{R}_c|} | b \rangle,
\end{equation}
where
\begin{equation}
 \langle a | \frac{-1}{|\mathbf{r} - \mathbf{R}_c|} | b \rangle=  \int \sigma_a(\mathbf{r}-\mathbf{R}_a) \frac{-1}{|\mathbf{r} - \mathbf{R}_c|} \sigma_b(\mathbf{r}-\mathbf{R}_b) d\mathbf{r}.
\end{equation}
Using the Gaussian product theorem, we define the new Gaussian centered in $\mathbf{R}_p$; the integral can then be written as:
\begin{equation}
 \langle a | \frac{-1}{|\mathbf{r} - \mathbf{R}_c|} | b \rangle=  -N_aN_b \exp\left(-\frac{\alpha_a \alpha_b}{\alpha_a + \alpha_b} |\mathbf{R}_a - \mathbf{R}_b|^2\right) \int  \frac{ \exp({-\alpha_p(|\mathbf{r}-\mathbf{R}_p)|^2})} {|\mathbf{r} - \mathbf{R}_c|}  d\mathbf{r}.
\end{equation}
It has been shown that
\begin{equation}
    \int  \frac{ \exp({-\alpha_p(|\mathbf{r}-\mathbf{R}_p)|^2})} {|\mathbf{r} - \mathbf{R}_c|}  d\mathbf{r}=\frac{2\pi}{\alpha_a + \alpha_b} F_0\left((\alpha_a + \alpha_b) |\mathbf{R}_p - \mathbf{R}_c|^2\right),
\end{equation}
with $F_0(t)$ being the zeroth order Boys function, which relates to the error function as:
\begin{equation}
F_0(t) = \frac{1}{2} \sqrt{\frac{\pi}{t}} \text{erf}(\sqrt{t}),
\end{equation}
with the interesting property of $\lim_{t\to 0} F_0(t) = 1$. The integral is then given by:
\begin{equation}
\langle a | \frac{-1}{|\mathbf{r} - \mathbf{R}_c|} | b \rangle = N_a N_b \frac{-2\pi}{\alpha_a + \alpha_b} \exp\left(-\frac{\alpha_a \alpha_b}{\alpha_a + \alpha_b} |\mathbf{R}_a - \mathbf{R}_b|^2\right) F_0\left((\alpha_a + \alpha_b) |\mathbf{R}_p - \mathbf{R}_c|^2\right).
\end{equation}
We can get $V_{pq}$ by summing over all contractions as shown in Eq. \eqref{V_pq}.

\subsection{Two-Electron Integral Computation Over Atomic Orbitals in the STO-3G basis}
The two-electron integral, as shown in Eq. \eqref{eq:h_pqrs-H2}, can be rewritten in AO basis as

\begin{equation}
h_{pqrs}= \sum_{\alpha \beta \gamma \delta=1}^2 c_{p\alpha} c_{q\beta} c_{r\gamma} c_{s\delta}  \int \int \phi_{\alpha}(\mathbf{r}_1) \phi_{\beta}(\mathbf{r}_1) \frac{1}{|\mathbf{r}_1 - \mathbf{r}_2|} \phi_{\gamma}(\mathbf{r}_2) \phi_{\delta}(\mathbf{r}_2) d\mathbf{r}_1 d\mathbf{r}_2,
\end{equation}
which, after developing the AO on the STO-3G basis, is written as:
\begin{equation}
\label{h_pqrs}
h_{pqrs} =
\sum_{\alpha \beta \gamma \delta=1}^2 c_{p\alpha} c_{q\beta} c_{r\gamma} c_{s\delta} \sum_{abcd=1}^{3} d_{\alpha a} d_{\beta b} d_{\gamma c} d_{\delta d} \langle ab|cd \rangle,
\end{equation}
where $\langle ab|cd \rangle$ is given by
\begin{equation}
\langle ab|cd \rangle = \int \int \sigma_a(\mathbf{r}_1- \mathbf{R}_a) \sigma_b(\mathbf{r}_1- \mathbf{R}_b) \frac{1}{|\mathbf{r}_1 - \mathbf{r}_2|} \sigma_c(\mathbf{r}_2- \mathbf{R}_c) \sigma_d(\mathbf{r}_2- \mathbf{R}_d) d\mathbf{r}_1 d\mathbf{r}_2,
\end{equation}
where $\sigma_a, \sigma_b, \sigma_c, \sigma_d$ are normalized primitive Gaussian functions defined above.\\
We use the Gaussian product theorem to reduce the two Gaussians on the left and the two Gaussians on the right; the new centers $\boldsymbol{R}_u$ and $\boldsymbol{R}_v$ are:
\begin{align}
\mathbf{R}_u &= \frac{\alpha_a \mathbf{R}_a + \alpha_b \mathbf{R}_b}{\alpha_a + \alpha_b},\\
\mathbf{R}_v &= \frac{\alpha_c \mathbf{R}_c + \alpha_d \mathbf{R}_d}{\alpha_c + \alpha_d}.
\end{align}
After further mathematical development, and using the Boys function, the integral takes this final form:
\begin{equation}
\langle ab|cd \rangle = N_a N_b N_c N_d \frac{(2\pi^2)^{\frac{5}{2}} \exp\left(-\frac{\alpha_a \alpha_b |\mathbf{R}_a - \mathbf{R}_b|^2}{\alpha_a + \alpha_b} - \frac{\alpha_c \alpha_d |\mathbf{R}_c - \mathbf{R}_d|^2}{\alpha_c + \alpha_d}\right)}{(\alpha_a + \alpha_b)(\alpha_c + \alpha_d)\sqrt{\alpha_a + \alpha_b + \alpha_c + \alpha_d}} F_0\left(\frac{(\alpha_a + \alpha_b)(\alpha_c + \alpha_d)}{\alpha_a + \alpha_b + \alpha_c + \alpha_d} |\mathbf{R}_u - \mathbf{R}_v|^2\right).
\end{equation}
We can get $h_{pqrs}$ by summing over all contractions as shown in Eq. \eqref{h_pqrs}.

\section{Basic VQE pipeline in Qiskit 1.2}
\label{app:vqe-pipeline}

The VQE, as outlined in Eq.\ref{eq:ritz}, can be broken down into several critical components, each of which requires careful decisions that influence the algorithm's structure and computational cost. This sequence of components is often referred to as the VQE pipeline. Decisions made regarding individual elements within this pipeline can have critical effects on the entire VQE procedure. In Figure \ref{fig:vqe-diagram}, we illustrate the iterative process, including the primary VQE loop, to provide a visual representation of the algorithm and its key components.

\subsection{Defining the Molecular Problem}
\label{app:sub:mol-problem}
The \texttt{Qiskit} SDK \cite{qiskit} and its ecosystem of companion packages greatly simplify the task of implementing a full VQE pipeline, starting from a geometric description of a molecule to ending with an estimation of its ground state energy on a QPU or a simulator. \texttt{Qiskit Nature} \cite{qiskit, qiskit_nature} provides an interface to the \texttt{PySCF} quantum chemistry library \cite{pyscf}, which is a Python toolkit that wraps C++ functions that perform Self-Consistent Field (SCF) method. Using \texttt{PySCF} through \texttt{Qiskit}, it is possible to define any molecule based on its constituent atoms and their spacial coordinates, in addition to the molecule's multiplicity or spin, as well as its charge to fix the number of electrons in the molecular system. It is also necessary to define the basis set's type and the unit of distance. These definitions are implemented by initializing a \texttt{PySCFDriver} from \texttt{Qiskit Nature}. The computations of the one- and two-body integrals for the second quantized Hamiltonian are performed internally by running the SCF algorithm to find the molecular orbitals and the Hartree-Fock reference state, then computing the integrals \eqref{eq:h_pq} and \eqref{eq:h_pqrs}. Running the \texttt{PySCFDriver} returns the electronic structure problem object, which contains the results of the computations mentioned above.\\

\begin{python}
# The BeH2 molecule
driver = PySCFDriver(
		    atom="""H -1.326, 0.0, 0.0
                Be 0.0, 0.0, 0.0
                H  1.326, 0.0, 0.0
             """,
		    basis='sto3g',
		    charge=0,
		    spin=0,
		    unit=UnitsType.ANGSTROM)

# Generating the Electronic Structure Problem
molecule_problem = driver.run()
\end{python}

We are also provided with a complete active space (CAS) method to focus on a specific set of active orbitals and freeze a set of occupied orbitals, which are, in general, the core orbitals. This reduces the number of required computational resources, notably the number of necessary qubits. This is performed using the \texttt{ActiveSpaceTransformer}, which reduces the original problem to a smaller problem. The transformer is instantiated by passing in the number of active electrons and the number of active molecular (spatial) orbitals. It is worth noting that both the full and reduced Hamiltonians we obtain either from the full problem or the reduced problem include constants that are computed classically, such as the nuclear-nuclear potential and all the residue terms that result from the active space reduction. These terms must be later re-introduced in the final result, as we will see in the upcoming subsections. Below, we reduce the entire problem to only consider $2$ electrons and $3$ active molecular orbitals, corresponding to a total of $6$ spin orbitals. These spin orbitals translate directly to qubits, first giving us $6$ qubits before we reduce this number further to $4$ qubits during the mapping step later.

\begin{python}
# Reducing the problem to the active space containing the 2 electrons in the 3 spatial orbitals
active_space_transformer = ActiveSpaceTransformer(
                        num_electrons=2, num_spatial_orbitals=3
                        )
reduced_molecule_problem = active_space_transformer.transform(molecule_problem)
\end{python}

\subsection{The Hamiltonian in terms of qubit operators}
Now that we have defined and reduced the molecular problem, one must generate the Hamiltonian and translate it into operators that can be directly measured on a quantum computer (spin or Pauli operators). This transformation, corresponding to a second quantization of the Hamiltonian and a mapping, can also affect both the depth of the ansatz and the required number of measurements. In \texttt{Qiskit}, obtaining the 2\tsp{nd} quantized Hamiltonian is a matter of extracting it from the problem object:

\begin{python}
# Obtaining the second quantized Hamiltonian
second_q_hamiltonian = reduced_molecule_problem.second_q_ops()[0]
\end{python}

As discussed in Sec.\ref{sec:mapping}, the next step is to map its ladder operators to Pauli operators. \texttt{Qiskit Nature} provides us with tools to perform this mapping, of which we will use the \texttt{ParityMapper}. The \texttt{ParityMapper} can apply a qubit tapering operation that reduces the number of qubits needed for the resulting mapped Hamiltonian if the number of electrons in the $\alpha$ and $\beta$ spin sectors is given. The \texttt{num\_particles} attribute of the molecule problem object gives these two numbers. The mapped Hamiltonian \texttt{qubit\_op} is now defined on $4$ qubits, as explained above.

\begin{python}
# Defining the Parity mapper
# When the number of particles is given, 2-qubit tapering is also applied
parity_mapper = ParityMapper(num_particles=reduced_molecule_problem.num_particles)
# Applying the Parity Mapping
qubit_op = parity_mapper.map(second_q_hamiltonian)
\end{python}

\subsection{Ansatz circuit construction in Qiskit}
The subsequent task involves selecting an ansatz that balances between computational expressiveness and practicality. It must be sufficiently expressive to approximate the ground state wave function accurately without leading to excessively deep circuits or overly complex parameterizations, making efficient training challenging.
Any parameterized circuit can, in principle, be used as an ansatz, provided it acts on the same number of qubits as the mapped Hamiltonian. These can be built manually or imported from \texttt{Qiskit} and \texttt{Qiskit Nature}'s circuit libraries. In our case, we shall use the ansätze provided to us in those libraries. As discussed in \ref{sec:vqe}, we will be using the \emph{UCCSD} and \emph{Efficient SU2} ansätze.

\subsubsection*{UCCSD}
We remind ourselves that in the \emph{UCCSD} ansatz, we typically evolve the Hartree-Fock initial state. We therefore must define this state for our reduced problem by passing in the number of electrons (in the $\alpha$ and $\beta$ sectors), the number of molecular orbitals, and the used mapper to the \texttt{HartreeFock} constructor.\\

\begin{python}
# Defining the Hartree-Fock initial
hf_initial_state = HartreeFock(num_particles=reduced_molecule_problem.num_particles,
                               num_spatial_orbitals=reduced_molecule_problem.num_spatial_orbitals,
                               qubit_mapper=parity_mapper)
\end{python}

We now build the \emph{UCCSD} circuit by passing to \texttt{UCCSD} the same parameters in addition to the initial state. The initial state will thus be prepended to the \emph{UCCSD} evolution circuit.\\

\begin{python}
# Defining the UCCSD ansatz using the HF initial state
ansatz = UCCSD(
        reduced_molecule_problem.num_spatial_orbitals,
        reduced_molecule_problem.num_particles,
        initial_state=hf_initial_state,
        qubit_mapper=parity_mapper
        )
\end{python}

\subsubsection*{Hardware-Efficient Ansatz: Efficient SU2}
\emph{HEA}s typically do not consider the physical properties of the system at hand, and such is the case for the \emph{Efficient SU2} ansatz. Therefore, its construction will mostly depend on the properties of the desired final quantum circuit, such as the number of qubits, the entanglement scheme, and the number of times the rotation and entanglement blocks are repeated. For a circuit with linear entanglement scheme, one repetition, and that acts on the same number of qubits as the mapped Hamiltonian, we used the definition below.\\

\begin{python}
ansatz = EfficientSU2(
            num_qubits=qubit_op.num_qubits, entanglement='linear', reps=1
            )
\end{python}

\subsection{Transpilation}
\label{app:vqe-pipeline.transpilation}
In order to run the ansatz quantum circuit on a quantum computer, we must re-express it in terms of quantum gates that are natively supported by the target quantum computer. That is, decomposing the initial logical quantum gates into native physical quantum gates, as well as respecting the physical qubits' connectivity which may require reassigning qubits and re-routing two-qubit gates. This process of converting a logical circuit into a physical one is called transpilation or compilation.\\\\
To implement this in \texttt{Qiskit}, first, we define the target quantum backend, which may be a real quantum hardware or a simulator. A variety of simulators can be used in \texttt{Qiskit}'s ecosystem, ranging from several perfect simulators to simulated IBM quantum computers. In our case, we use \texttt{Qiskit Aer}'s \texttt{AerSimulator} \cite{qiskit}. Then, we prepare a pass manager to perform the transpilation. We set the target backend to the simulator and specify that we require no optimization in our specific case. The transpiled quantum circuit is the Instruction Set Architecture (ISA) circuit. Finally, the qubit layout of the transpiled ansatz is applied to the Hamiltonian observable, re-routing the qubits of the observable to align with those of the ansatz. This is important as we must measure the correct observables on the correct qubits.\\

\begin{python}
# Creating a backend
# In this case, it is an SVS simulator
backend = AerSimulator()
# Creating the pass manager that transpiles the ansatz
pm = generate_preset_pass_manager(backend=backend, optimization_level=0)

# Transpiling the ansatz
# ISA stands for "Instruction Set Architecture"
isa_ansatz = pm.run(ansatz)

# Applying the layout of the ISA ansatz to the Hamiltonian observable
# This ensures that the observables qubits are the same as the ansatz qubits
isa_observables = qubit_op.apply_layout(isa_ansatz.layout)
\end{python}

\subsection{Measuring eigenvalues and energies}
\texttt{Qiskit} uses the \emph{primitives} processing instructions \cite{IBMQ2024} in order to interact with real quantum hardware. They are defined as the simplest building blocks of quantum applications. Two primitives are available: \texttt{Sampler} and \texttt{Estimator}. The former is used to directly measure the qubits' states, whereas the latter is used in addition to a set of observables to measure their expectation values with respect to the state defined by an input quantum circuit. For the VQE, we must use the \texttt{Estimator} primitive to measure the Hamiltonian's expectation value with respect to our parameterized ansatz.\\
Let us remind ourselves that the reduced Hamiltonian we have defined above omits certain constant terms that are stored in the molecule problem's instance, as mentioned in \ref{app:sub:mol-problem}. Therefore, these constants must be added back to the expectation values that we measure using \texttt{Estimator}. At the end, given a parameterized ansatz $\ket{\psi}$, a Hamiltonian $H$, and a set of variational parameters $\vtheta$, the sum of these omitted constants and the resulting expectation value gives us the value of the energy cost function $E(\vtheta) = \bra{\psi(\vtheta)} H \ket{\psi(\vtheta)}$.\\
The two code blocks below define the expectation value correction (interpretation) and the energy cost functions.\\

\begin{python}
# Getting the energy value by interpreting the expectation value
# in the context of the reduced molecule problem
def interpret_exp_val(exp_val, problem):
    # Wrapping the expectation value in MinimumEigensolverResult
    sol = MinimumEigensolverResult()
    sol.eigenvalue = np.real(exp_val)
    # Interpreting the result
    return  problem.interpret(sol).total_energies[0]
\end{python}

\begin{python}
# Using the Estimator primitive with the previously defined backend
estimator = EstimatorV2(mode=backend)

# Defining the energy cost function
def energy_cost_function(params):
    # Run the job and get the eigenvalue result
    estimator_job = estimator.run([(isa_ansatz, isa_observables, params)])
    estimator_exp_val = estimator_job.result()[0].data.evs
    # Return the interpreted energy value
    return interpret_exp_val(estimator_exp_val, reduced_molecule_problem)
\end{python}

\subsection{The optimization}
The core classical component of the VQE is the optimization algorithm. It is the classical procedure of varying a set of parameters to minimize the value of the cost function. Of the various optimizer algorithms that exist \cite{powell1964, powell1998, sorourifar2024}, the Simultaneous Perturbation Stochastic Approximation (SPSA) is one of the most adequate optimizers for VQE applications, as it is designed for fluctuating cost functions \cite{spall1998}, and performs well under noisy conditions \cite{pellow2021}. We will make use of SPSA to apply the variational principle in Eq.\ref{eq:ritz} until the expectation value for the Hamiltonian is minimized. A description of SPSA's workflow is given in subsection \ref{subsec:optimization}.\\\\
\texttt{Qiskit}'s \texttt{SPSA} optimizer takes in the cost function to minimize as well as an initial set of parameters. During the optimization, these parameters will be varied to minimize the value of the cost function. This will go on until a set number of iterations is reached or until a predefined termination condition is met. For our VQE, the cost function is the energy cost function, and the parameters are the ansatz parameters. We chose to limit the SPSA optimization to $250$ iterations instead of setting a termination condition. This is implemented in the code below, where we also define a callback function that is run after every iteration. It stores the intermediate energy values and prints them while the VQE is ongoing.

\begin{python}
# Results list to store the energy values
results = []

# The callback function runs after each iteration
def optimizer_callback(ne, params, value, step, accepted):
    global results
    results.append(value)
    print(f'Iteration {len(results):03d}  -  Energy = {value}')
\end{python}

\begin{python}
# Defining the SPSA optimizer
optimizer = SPSA(maxiter=250, callback=optimizer_callback)
\end{python}

\subsubsection*{Calibration of the learning rate}
The SPSA algorithm calibrates its learning rate using a first set of $50$ measurements to estimate the gradient around the initial point in the search space defined by the initial parameters \cite{spall2003, kandala2017}. The remaining hyperparameters of \texttt{SPSA} are left to their \texttt{Qiskit} implementation's default values. (see subsection \ref{subsec:sim}.)

\begin{python}
# Defining random initial parameters
initial_params = np.random.rand(ansatz.num_parameters)

# Calibrating the SPSA optimizer
learning, pert = optimizer.calibrate(energy_cost_function, initial_params)

# Setting the calibrated learning rate and perturbation series
optimizer.learning_rate = learning
optimizer.perturbation = pert
\end{python}

\subsection{Running the VQE}
At this point, every component of the VQE is defined and set: the molecule's Hamiltonian, the ansatz, the energy cost function, and the optimization algorithm. The VQE can then be performed by running the optimizer. The final result will contain the optimized ansatz parameters and the minimized energy value. If the optimizer reaches the global minimum of the cost function, then the minimized energy will correspond to the molecule's ground state energy.

\begin{python}
# Running the VQE algorithm
# It is the SPSA algorithm that uses a quantum cost function
result = optimizer.minimize(energy_cost_function, initial_params)

# Storing the final energy and parameters
energy_result, parameters_result = result.fun, result.x
\end{python}

\subsection{VQE results and post-processing}
In the VQE, we are not always guaranteed to reach a good result due to the possibility of encountering local minima, barren plateaus, or simply as a result of excessive noise. In the cases where the algorithm doesn't converge towards a meaningful value, we have to restart the optimization or make any necessary changes in any of the components of the VQE's pipeline.
Provided that a satisfactory result - in the limits of quantum coherent and incoherent noise - is reached, error mitigation techniques should, in general, be applied \cite{Tilly2022, dobrautz2024}. Several of these have been developed for the NISQ era, such as zero-noise extrapolation (ZNE), Pauli Twirling, Clifford data regression, probabilistic error amplification, and probabilistic error cancellation to name a few. In the following, we provide a brief description of ZNE, as well as a simple implementation in \texttt{Qiskit 1.2}.

\subsubsection{Zero-Noise Extrapolation}
ZNE \cite{temme2017} is an error mitigation strategy for NISQ-era quantum computing aimed at enhancing the accuracy of expectation values measurement on noisy quantum systems. The fundamental principle of ZNE involves deliberately increasing the noise levels of quantum circuits, often achieved by techniques like stretching gate durations or amplifying gate errors by repeating the gates or the whole circuit and then using the new resulting noisier data points to extrapolate back towards an estimate of what the outcome would be without noise. As an error \emph{mitigation} technique, this approach avoids the need for additional quantum resources for complex error-correcting codes at the cost of increasing the number of quantum computations and additional classical post-processing. This makes it particularly suitable for near-term quantum devices. By mitigating the effects of noise through this extrapolation, ZNE can potentially improve the accuracy of quantum algorithms such as VQE, helping to bridge the gap between current noisy hardware and accurate quantum simulations.
In the code below, We implement a simple ZNE for a fake QPU. The fake QPU (namely, a noisy simulator) replaces the ideal simulator we have used in \ref{app:vqe-pipeline.transpilation}.

\begin{python}
# Since we will produce new circuits, we need another transpiler.
zne_pm = generate_preset_pass_manager(
                     optimization_level=0,
                     basis_gates=backend.configuration().basis_gates
                     )

def fold(isa_circuit, n=1):
    """Creates integer folds for a given transpiled circuit"""
    new_circuit = isa_circuit.copy()

    for _ in range((n-1)//2):
        new_circuit = new_circuit.compose(isa_circuit.compose(isa_circuit.inverse()))

    # The new circuit must re-transpiled
    new_circuit = zne_pm.run(new_circuit)
    return new_circuit
\end{python}

Other methods of post-processing include, for example, averaging over the last $N$ iterations' results \cite{kandala2017} to get a final estimate of the Hamiltonian's ground state energy.

\end{document}